\documentclass[aps,pra,twocolumn,showpacs,superscriptaddress]{revtex4-1}
\usepackage[colorlinks,linkcolor=blue,anchorcolor=blue,citecolor=blue]{hyperref}
\usepackage{graphicx,amsmath,amssymb}
\usepackage[usenames]{color}
\bibpunct{[}{]}{,}{s}{}{;}

\begin{document}

\title{Nodes and Spin Windings for Topological Transitions in Light-Matter
Interactions: \\
Anisotropic Quantum Rabi Model as a Born Abstract Artist
}
\author{Zu-Jian Ying }
\email{yingzj@lzu.edu.cn}
\affiliation{School of Physical Science and Technology,
Lanzhou University, Lanzhou 730000, China}

\begin{abstract}
By extracting different levels of topological information a new light is shed on the energy spectrum of
the anisotropic quantum Rabi model (QRM) which is the fundamental model of
light-matter interactions with indispensable counter-rotating terms in
ultra-strong couplings. Besides conventional topological transitions (TTs) at gap closing,
abundant unconventional TTs including a particular one universal for different energy levels are unveiled underlying level
anticrossings without gap closing by tracking the wave-function nodes.
On the other hand, it is found that the nodes have a correspondence to spin
windings, which not only endows the nodes a more explicit topological
character in supporting single-qubit TTs but also turns the topological information physically detectable.
Furthermore, hidden small-spin-knot transitions are exposed for the ground state, while more kinds of spin-knot
transitions emerge in excited states including unmatched node numbers and spin winding numbers.
As a surprise, frequently the spin windings produce portraits in high spiritual similarity with abstract artistic works,
which demonstrates that the anisotropic QRM may be the Picasso of physical models. This signifies that art is joining the dialogue
between mathematics and physics which was triggered by the milestone work of revealing integrability of the QRM.
\end{abstract}
\pacs{ }
\maketitle


\section{Introduction}

Among the intensive dialogue between mathematics and physics
\cite%
{Braak2011,Solano2011,Boite2020,Liu2021AQT,Diaz2019RevModPhy,Kockum2019NRP,Rabi-Braak,Braak2019Symmetry,Wolf2012,FelicettiPRL2020,Felicetti2018-mixed-TPP-SPP,Felicetti2015-TwoPhotonProcess,Simone2018, Irish2014,Irish2017,Irish-class-quan-corresp,
PRX-Xie-Anistropy,Batchelor2015,XieQ-2017JPA,
Hwang2015PRL,Bera2014Polaron,Hwang2016PRL,Ying2015,LiuM2017PRL,Ying-2018-arxiv,Ying-2021-AQT,Ying-gapped-top,Ying-Stark-top,Grimaudo2022q2QPT,
CongLei2017,CongLei2019,Ying2020-nonlinear-bias,ChenQH2012,
e-collpase-Garbe-2017,e-collpase-Duan-2016,Garbe2020,Rico2020,e-collpase-Garbe-2017,
Garbe2020,Garbe2021-Metrology,Ilias2022-Metrology,Ying2022-Metrology,
Boite2016-Photon-Blockade,Ridolfo2012-Photon-Blockade,Li2020conical,
Ma2020Nonlinear,
ZhangYY2016,ZhengHang2017,PengJie2019,Liu2015,Ashhab2013, ChenGang2012,FengMang2013,Eckle-2017JPA,Casanova2018npj,HiddenSymMangazeev2021,HiddenSymLi2021,HiddenSymBustos2021,Garbe2020,
JC-Larson2021,Stark-Cong2020,Cong2022Peter,Stark-Grimsmo2013}
triggered by the milestone work of D. Braak who
revealed the integrability of the quantum Rabi model (QRM),~\cite{Braak2011}
few-body quantum phase transitions (QPTs) have recently attracted a special
attention in the context of light-matter interactions. \cite%
{Liu2021AQT,Ashhab2013,Ying2015,Hwang2015PRL,Ying2020-nonlinear-bias,Ying-2021-AQT,LiuM2017PRL,Hwang2016PRL,Ying-gapped-top,Ying-Stark-top,Ying-2018-arxiv,Grimaudo2022q2QPT}
In reality, the continuing experimental enhancements of couplings have
brought the era of ultra-strong~\cite%
{Diaz2019RevModPhy,Kockum2019NRP,Wallraff2004,Gunter2009,Niemczyk2010,Peropadre2010,FornDiaz2017,Forn-Diaz2010,Scalari2012,Xiang2013,Yoshihara2017NatPhys,Kockum2017}
and even deep-strong couplings,~\cite{Yoshihara2017NatPhys,Bayer2017DeepStrong}
which makes few-body QPTs practically relevant. Along with the applications
of few-body QPTs e.g. in critical quantum metrology,~\cite%
{Garbe2020,Garbe2021-Metrology,Ilias2022-Metrology,Ying2022-Metrology}
single-qubit topological phase transitions have also added an interesting
topic in the mathematics-physics dialogue with a renewed insight for the
transitions in light-matter interactions.~\cite%
{Ying-2021-AQT,Ying-gapped-top,Ying-Stark-top}

In the QRM,~\cite{rabi1936,Rabi-Braak} which is a most fundamental model of
light-matter interactions, the QPT \cite{Ashhab2013,Ying2015,Hwang2015PRL}
occurs in the low frequency limit, that is, $\omega /\Omega \rightarrow 0$
where $\omega $ is the bosonic frequency and $\Omega $ is the atomic level
splitting or tunneling strength, which is a replacement of thermodynamical
limit in condensed matter. Although whether the transition should be termed
quantum or not might be a matter of taste due to the negligible quantum
fluctuations in the photon vacuum state,\cite{Irish2017} it has been
established that the critical exponents of the single-qubit QRM can be
bridged to the thermodynamical case. \cite{LiuM2017PRL} Indeed, the photon
number has a superradiant-like behavior~\cite%
{Ashhab2013,Ying2015,Ying-2021-AQT,Ying-Stark-top} as experimentally
observed~\cite%
{SuperrianceDickeTheo1,SuperrianceDickeTheo2,SuperrianceDickeExpr} in the
Dicke model which is a thermodynamical version of the QRM. Moreover, when
universality is a character born with QPT,~\cite{Sachdev-QPT,Irish2017} the
QPT of QRM manifests a universal critical scaling relation with respect to
the anisotropy of linear coupling,~\cite%
{LiuM2017PRL,Ying-2021-AQT,Ying-Stark-top} and more robust scaling relations
can be found in the presence of nonlinear coupling.~\cite{Ying-Stark-top}
Despite of the preserved parity symmetry in the Hamiltonian of the QRM, such a
QPT has a hidden symmetry breaking~\cite{Ying-2021-AQT} in the ground state
as in the traditional Landau class of phase transitions.

Away from the low frequency limit, the afore-mentioned critical universality
however collapses at finite frequencies.~\cite{Ying-2021-AQT,Ying-Stark-top}
Surprisingly, in such a diversified situation~\cite{Footnote-Diversity}
another universality classification can be found from the common node
numbers of the ground-state wave functions, which reveals the topological
nature of the emerging series of transitions with symmetry protection,
essentially different from the Landau class of phase transitions.~\cite%
{Ying-2021-AQT,Ying-gapped-top,Ying-Stark-top} Such topological transitions
(TTs) are usually conventional ones with gap closing as those in condensed
matter,~\cite%
{Topo-Wen,Hasan2010-RMP-topo,Yu2010ScienceHall,Chen2019GapClosing,TopCriterion,Top-Guan,TopNori}
while unconventional ones without gap closing also exist~\cite%
{Ying-gapped-top} analogously to the unconventional cases in the quantum
spin Hall effect with strong electron-electron interactions~\cite%
{Amaricci-2015-no-gap-closing} and the quantum anomalous Hall effect with
disorder.~\cite{Xie-QAH-2021} The topological classification applies not
only for linear couplings but also for the nonlinear Stark interaction.~\cite%
{Ying-Stark-top} While nodes of wave functions are the center of Feynman's
node theorem which generally governs one-dimensional confined spinless systems,~\cite%
{Ref-No-node-theorem} it is demonstrated that the ground state of the QRM
also obeys the no-node theorem.~\cite{Ying-gapped-top} Nodes of polynomial
functions are related to topological Galois theory in connecting algebra to
topology.~\cite{topo-Galois-theory} The node number can be used to
distinguish topological difference of quantum states in the sense that by
fixing a node number one cannot go to another node state by continuous shape
deformation of the wave function, just as one cannot change a torus into a
sphere by a continuous deformation. Besides such a topological picture as in
so-called rubber-sheet geometry, one may be wondering
if there is any more physical demonstration for the topological connection
of nodes in wave functions.

On the other hand, for the light-matter interaction it has been shown that
it is necessary to take the counter-rotating terms (CRTs) into account~\cite%
{PRX-Xie-Anistropy} to fit the experimental energy spectrum in ultra-strong
couplings.~\cite{Forn-Diaz2010} In fact, the anisotropy describing the CRTs
is an important parameter in experiments which is highly controllable.~\cite%
{Yimin2018} In the absence of the CRTs the system denoted by the
Jaynes-Cummings model (JCM)~\cite{JC-model} possesses symmetries of the
excitation number and momentum-position duality.~\cite%
{Braak2019Symmetry,Ying-2021-AQT} The CRTs break both the symmetries
including the U(1) symmetry so that the excitation number is not longer a
good quantum number, while the emerging level anticrossings (also called
avoided crossings) lead to a different energy spectrum. Although the $Z_2$
symmetry of parity is preserved in the presence of the CRTs, the parity is
not enough at all to label the various quantum states. It has been shown
that level crossings occur on the so-called ``baselines'\ in coupling
variation,~\cite{Braak2019Symmetry} while it is not clear what is really
happening to the quantum states during level crossings and anticrossings.
The afore-mentioned node number has been applied for a renewed
classification of the ground state,~\cite%
{Ying-2021-AQT,Ying-gapped-top,Ying-Stark-top} however it is still
unexplored for the excited states which may have different scenarios. In
such a situation, a full identification and a complete understanding for all
quantum states in the presence of the CRTs are still lacking even when the
QRM has been long studied for over eight decades.~\cite{rabi1936,Rabi-Braak}

Conventionally in condensed matter topological phases are concerning about
the ground state,~\cite{ColloqTopoWen2010} while for the ground state
single-qubit systems in light-matter interactions also have analogs of TTs.~%
\cite{Ying-2021-AQT,Ying-gapped-top,Ying-Stark-top} Since state occupation
of single-qubit systems only involves a single eigenstate which is free of
filling problem in condensed matter with Fermi level, to get some
topological insight for the energy spectrum we can extend the exploration of
TTs to excited states. In the present work, we provide a topological point
of view for the energy spectrum of the anisotropic QRM. We find that not
only the level crossings but also the level anticrossings are associated
with TTs as tracked by the node number of the eigen-state wave function. On
the other hand, we show that the nodes of the wave function corresponds to
the zeros of spin windings, which endows nodes an explicit topological
character and provides a more solid support for single-qubit TTs. In such a
topological classification, the spin winding direction contributes an
additional quantum feature for a full identification of various ground
states. For the excited states, apart from the TTs associated with level
crossing or anticrossing, novel TTs with unmatched node numbers and winding
numbers also emerge due to anti-winding large spin knots. A more extensive
investigation on spin knots reveals hidden small-knot transitions for the
ground state and more kinds of spin-knot transitions for the excited states.
Our study extracts different levels of topological information, from
wave-function topology to spin-winding topology, with exploration from
one-axis nodes to two-axis nodes and in the absence or presence of various
spin knots including small, big, huge and diagonal ones. The abundant
topological information may renew our knowledge about the underlying
variation of the quantum states in light-matter interactions. Besides the
unveiled variety of novel TTs, a surprising finding is that the figures
produced by the spin windings frequently resemble various portraits as in
abstract artistic works in such an amazing spiritual similarity that we
believe the anisotropic QRM is a born abstract artist.

The paper is organized as follows. Section \ref{Sect-Model} introduces the
anisotropic QRM which is a fundamental model describing light-matter
interactions with the CRTs. Section \ref{Sect-Spectrum} illustrates the
energy spectrum with level crossing and anticrossings. Section \ref%
{Sect-Nodes-Anticrossings} unveils underlying TTs of node numbers around
the anticrossings and an unconventional TT universal for different states.
The correspondence of nodes and spin windings is shown in
Section \ref{Sect-Spin-Windings}, which yields a full topological
classification including winding directions for the ground state. Section %
\ref{Sect-Bridge-Knot-TopTrans} reveals new TTs with unmatched node numbers
and winding numbers in the excited states. Section \ref%
{Sect-Node-Sort-Algebraic-Formul} gives node sorting and algebraic formation
for spin winding number, which inspires and facilitates the exploration of
the transitions of different spin knots in Section \ref%
{Sect-Hidden-Knots-Trans}. Section \ref{Sect-Conclusions} is devoted to
conclusions and discussions.

\section{Model and Symmetry}

\label{Sect-Model}

The continuing experimental progresses in enhancing the strength of
light-matter interactions
have brought us to the era of ultra-strong\cite%
{Diaz2019RevModPhy,Wallraff2004,Gunter2009,Niemczyk2010,Peropadre2010,FornDiaz2017, Forn-Diaz2010,Scalari2012,Xiang2013,Yoshihara2017NatPhys,Kockum2017}
and even deep-strong couplings, \cite%
{Yoshihara2017NatPhys,Bayer2017DeepStrong} with the ratio of the coupling
strength and the bosonic frequency going beyond $0.1$ and $1.0$
respectively. In such coupling regimes, the CRTs play an indispensable role,
as shown by the Bloch-Siegert shift \cite{Forn-Diaz2010,Pietikainen2017} and
the fitting \cite{PRX-Xie-Anistropy} of the experimental spectra.\cite%
{Forn-Diaz2010} Indeed the CRTs are controlled by the coupling anisotropy
ratio $\lambda $, while recently a very wide strength range of anisotropy
ratios from $\lambda =0.2$ to $\lambda =2.88$ have been experimentally tuned
to.\cite{Yimin2018} Access to ultra-strong couplings can be also possible for $\lambda =0$ in circuit-QED systems.\cite{Ulstrong-JC-1,Ulstrong-JC-2} In such a situation, the anisotropic coupling is
described by the anisotropic QRM of which the Hamiltonian reads
\begin{equation}
H=\omega a^{\dagger }a+\frac{\Omega }{2}\sigma _{x}+g\left[ \left(
\widetilde{\sigma }_{-}a^{\dagger }+\widetilde{\sigma }_{+}a\right) +\lambda
\left( \widetilde{\sigma }_{+}a^{\dagger }+\widetilde{\sigma }_{-}a\right) %
\right] .
\end{equation}%
Here $\omega $ is the frequency of a bosonic mode created (annihilated) by $%
a^{\dagger }$ ($a$) and $\sigma _{x,y,z}$ are the Pauli matrices. The
anisotropy ratio $\lambda $ tunes the afore-mentioned CRTs which, along with
the rotating-wave terms, contribute to the total coupling with strength $g$.
Following refs.\cite{Irish2014,Ying2015} we have used the spin basis $\sigma
_{z}=\pm $ which conveniently represents the two flux states in the
flux-qubit circuit systems,\cite{flux-qubit-Mooij-1999} with an
unconventional definition of spin raising and lowering operators $\widetilde{%
\sigma }^{\pm }=(\sigma _{z}\mp i\sigma _{y})/2$. Nevertheless, the
conventional form of the QRM ($\lambda =1$) can be retrieved by a spin
rotation \{$\sigma _{x},\sigma _{y},\sigma _{z}$\} $\rightarrow $ \{$\sigma
_{z},-\sigma _{y},\sigma _{x}$\} around the axis $\vec{x}+\vec{z}$. The $%
\Omega $ term denotes the atomic level splitting in cavity systems and
tunneling \cite{Ying2015,Irish2014} in flux-qubit circuit systems.

We transform into the effective spatial space%
\begin{equation}
H=\frac{\omega }{2}\hat{p}^{2}+v_{\sigma _{z}}(x)+[\frac{\Omega }{2}-g_{y}i%
\sqrt{2}\hat{p}]\sigma ^{+}+[\frac{\Omega }{2}+g_{y}i\sqrt{2}\hat{p}]\sigma
^{-}  \label{Hx}
\end{equation}%
by the quadrature representation, $a^{\dagger }=(\hat{x}-i\hat{p})/\sqrt{2},$
$a=(\hat{x}+i\hat{p})/\sqrt{2}$, with momentum $\hat{p}=-i\frac{\partial }{%
\partial x}$. Now $\sigma _{x}=\sigma ^{+}+\sigma ^{-}$, $\sigma
_{y}=-i(\sigma _{+}-\sigma _{-})$ are the conventional forms of spin raising
and lowering on the $\sigma _{z}=\pm $ basis. The effective harmonic
potentials $v_{\sigma _{z}}(x)=\omega \left( x+g_{z}^{\prime }\sigma
_{z}\right) ^{2}/2+\varepsilon _{0}^{z}$ include a spin-dependent
displacement $g_{y,z}^{\prime }=\sqrt{2}g_{y,z}/\omega $, where $g_{y}=\frac{%
\left( 1-\lambda \right) }{2}g$ and $g_{z}=\frac{\left( 1+\lambda \right) }{2%
}g$, and a constant shift $\varepsilon _{0}^{z}=-\frac{1}{2}[g_{z}^{\prime
2}+1]\omega $. The $g_{y}$ term can be written as $\sqrt{2}g_{y}\hat{p}%
\sigma _{y}$ which resembles the Rashba spin-orbit coupling in nanowires
\cite{Nagasawa2013Rings,Ying2016Ellipse,Ying2017EllipseSC,Ying2020PRR,Gentile2022NatElec}
or the equal-weight mixture \cite%
{LinRashbaBECExp2011,LinRashbaBECExp2013Review} of the linear Dresselhaus ($%
\hat{p}_{x}\sigma _{y}+\hat{p}_{y}\sigma _{x}$) and Rashba ($\hat{p}%
_{x}\sigma _{y}-\hat{p}_{y}\sigma _{x}$) spin-orbit couplings in condensed
matter\cite{Dresselhaus1955,Rashba1984} and cold atomic gases.\cite%
{LinRashbaBECExp2011,LinRashbaBECExp2013Review,Li2012PRL} We will analyze by
standing in positive-$\lambda $ regime and $x$ space while the picture for
negative-$\lambda $ regime is similar in $p$ space under the $x$-$p$ duality
transformation $\left\{ \sigma _{x},\sigma _{y},\sigma _{z}\right\}
\rightarrow \left\{ \sigma _{x},-\sigma _{z},\sigma _{y}\right\}
,x\rightarrow p,\lambda \rightarrow -\lambda $.~\cite%
{Ying-2021-AQT,Ying-gapped-top,Ying-Stark-top}

The model at any anisotropy has the parity symmetry, $[\hat{P},H]=0$ with $%
\hat{P}=\sigma _{x}(-1)^{a^{\dagger }a}$ which includes simultaneously the
spin reversion and the space inversion $x\rightarrow -x$.\cite%
{Ying2020-nonlinear-bias,Ying-2021-AQT} The parity not only impose the
symmetry on the wave function but also leads to symmetric/antisymmetric spin
textures as addressed later on.

\section{Level Anti-Crossings in Energy spectrum}

\label{Sect-Spectrum}

\begin{figure}[t]
\includegraphics[width=1.0\columnwidth]{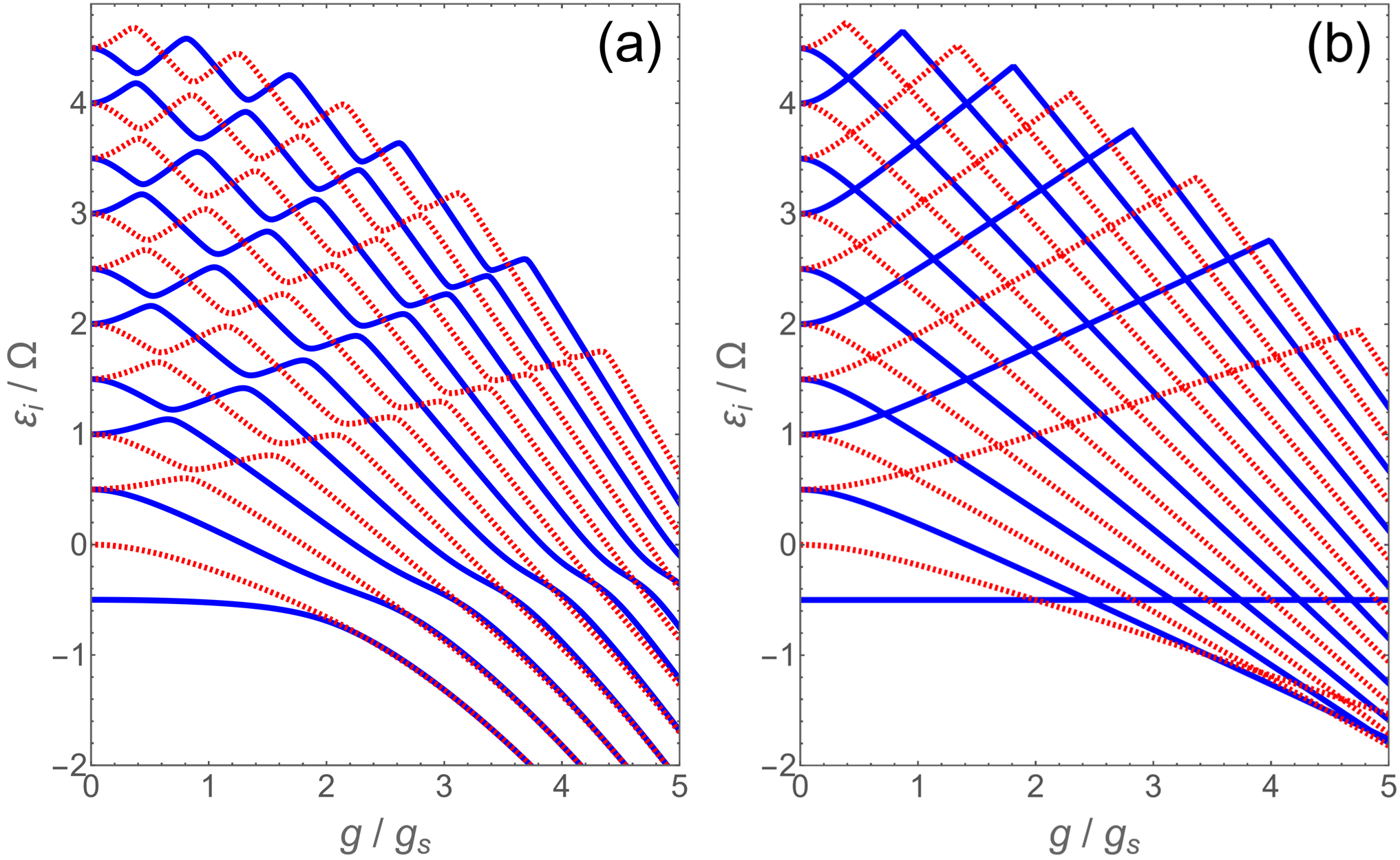}
\caption{Level crossing and anticrossings in the energy spectrum $\varepsilon _i$ at fixed
anisotropy: (a) $\protect\lambda=0.5$, (b) $\protect\lambda=0.0$. Here $i=j_E=1,2,\ldots$ labels the energy levels from the ground state to excited states. The blue solid (red dotted) lines represent
levels in positive (negative) parity. All figures are illustrated at $\omega=0.5\Omega$ throughout the paper. }
\label{fig-Ej}
\end{figure}

An illustration of the energy spectrum is presented in \textbf{Figure} \ref%
{fig-Ej}, where the levels with a negative (positive) parity are plotted by
blue solid (red dotted) lines. One sees in Figure \ref{fig-Ej}a that in the
presence of the CRTs ($\lambda\neq 0$) the energy levels exhibit level
crossings between states with opposite parities, while level anticrossings
occur between states with same parities. In contrast, there is no
anticrossing in the absence of the CRTs ($\lambda = 0$) as in Figure \ref%
{fig-Ej}b.

The full level crossings at $\lambda =0$ are results of the symmetries of
the JCM which preserves the parity, the excitation number and the
position-momentum duality.~\cite%
{Braak2019Symmetry,Ying-2021-AQT,Ying-gapped-top} However, the symmetries of
the excitation number and the position-momentum duality are broken once the
CRTs are introduced, which is the origin of the level anticrossings.
Nevertheless the symmetry of parity is still preserved, which partially
keeps level crossings. In fact, both the JCM part and the CRTs commute with
the parity operator. Around the level anticrossings the CRTs mix the
crossing states with a same parity in the absence of the CRTs, which opens a
gap. However, around the level crossings such a state mixing is excluded as
any combination of states with opposite parities would break the parity
symmetry thus violating the conditions for eigenstates.

Despite that the energy spectrum is known, one can never say the physics
behind has been fully explored. In the following we will provide some novel
insights for the level crossings and anticrossings from topological point of
view.

\section{Node Feature for Topological Classification}

\label{Sect-Nodes-Anticrossings}

\subsection{Transitions of Wave-Function Node Status in Level Crossings and
Anticrossings}

To find out the possible transitions of quantum states around the level
crossings and anticrossings, we monitor the wave function in continuous
variation of a system parameter. As mentioned in Introduction, in the
low-frequency limit the ground state of the generalized QRM with anisotropy~%
\cite{LiuM2017PRL,Ying-2021-AQT,Ying-gapped-top,Ying-Stark-top} and
nonlinear Stark coupling~\cite{Ying-Stark-top} can exhibit critical
universality which however breaks down at finite frequencies.~\cite%
{Ying-2021-AQT,Ying-Stark-top} Nevertheless, in the breaking down of the critical
universality another class of universality, namely topological universality,
can be found from the common feature of node number of the ground-state wave
function.~\cite{Ying-2021-AQT,Ying-gapped-top,Ying-Stark-top} As one knows a
common example of a topological invariant is the number of holes in an
object, here the topological invariant is the number of nodes in a
ground-state wave function. We find such topological classification also
applies for excited states and can provide a novel view for the level
crossings and anticrossings.

\begin{figure}[t]
\includegraphics[width=1.0\columnwidth]{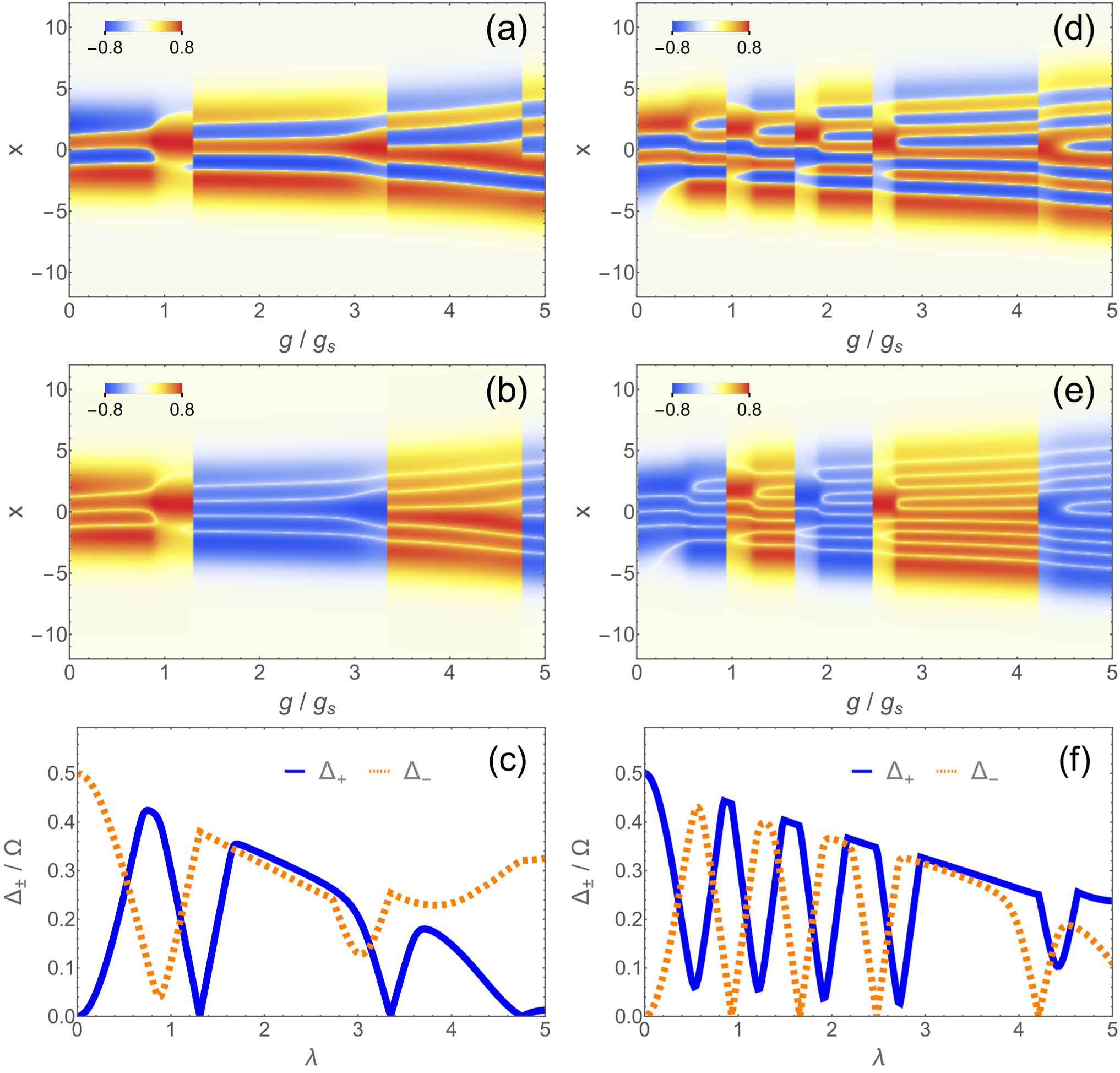}
\caption{Connection of level crossing and anticrossing with topological
transitions in variation of coupling $g$, at fixed $\protect\lambda=0.2
\Omega$ for $j_E=5$ (a-c) and fixed $\protect\lambda=0.3 \Omega$ for $j_E=10$
(d-f). (a,d) $\protect\psi_+(x)$, (b,e) $\protect\psi_+^*(x)\times \protect%
\psi_-(-x)$, (c,f) $\Delta_\pm$. The plot amplitude is amplified e.g. by $|\psi_{\pm}|^{1/4}$ to increase the color contrast. }
\label{fig-Nodes-Crossings-fix-Lambda}
\end{figure}

The nodes are defined as the zeros of wave function which are decided by
\begin{equation}
\psi _{+}\left( x_{Z}\right) =P\psi _{-}\left( -x_{Z}\right) =0,
\label{def-nodes}
\end{equation}
under parity value $P=\pm 1$, while the node number $n_Z$ counts the node
pairs $\pm x_{Z}$ totally in $\psi_ {\pm}$. \textbf{Figure}~\ref%
{fig-Nodes-Crossings-fix-Lambda}a,d show the wave function in spin-up
component $\psi _+(x)$ versus the increase of the coupling $g$, with blue
(red) color representing negative (positive) values, while the nodes appear
as the red/blue boundaries. Figure~\ref{fig-Nodes-Crossings-fix-Lambda}b,e
denote the product of the wave-function components,
\begin{equation}
\psi _+^*(x) \psi _-(-x)=P|\psi _+(x)|^2,  \label{WaveProduct}
\end{equation}
which indicates a negative (positive) parity in blue (red) while the track
of a node appears as a white line. Figure~\ref%
{fig-Nodes-Crossings-fix-Lambda}c,f depict the upper gap $\Delta _{+}$
(lower gap $\Delta _{-}$) by the blue solid (orange dotted) lines, with
vanishing value marking a level crossing and finite dip displaying an
anticrossing.

\begin{figure}[t]
\includegraphics[width=1.0%
\columnwidth]{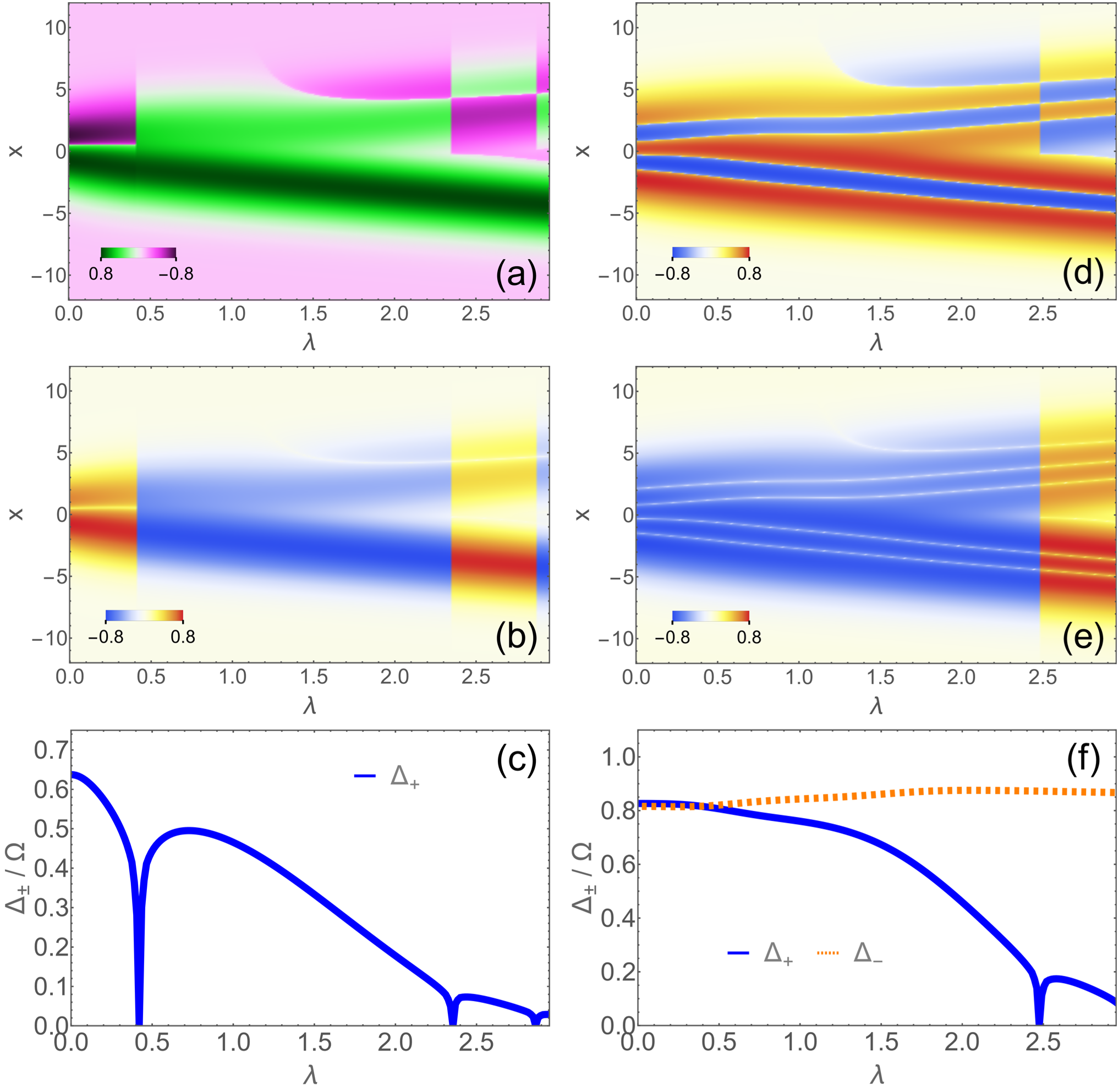}
\caption{An unconventional TT without gap closing universal for different
eigenstates beyond $\protect\lambda=1$, illustrated for $j_E=1$ (a-c) and $%
j_E=5$ (d-f) at fixed $g= 2.2\Omega $ and $\protect\omega=0.5\Omega$. (a,d) $%
\protect\psi_+(x)$, (b,e) $\protect\psi_+^*(x)\times \protect\psi_-(-x)$,
(c,f) $\Delta_\pm$. }
\label{fig-Nodes-Crossings-fix-g}
\end{figure}

Figure~\ref{fig-Nodes-Crossings-fix-Lambda}a-c illustrate the case for the
fifth excited states ($j_E=5$). As expected, the level crossings are always
accompanied with transitions of both parity and node number, as the
transitions around $g=1.3g_{\mathrm{s}}$ and $g=3.4g_{\mathrm{s}}$. The
situation is different for a level anticrossing as there is no parity
reversal while there are still transitions of node number for a small
anticrossing gap, as in the example around $g=0.9g_{\mathrm{s}}$ in Figure~%
\ref{fig-Nodes-Crossings-fix-Lambda}a-c. For a larger anticrossing gap, as
in the case around $g=3g_{\mathrm{s}}$, one can still see the tendency of
node modification despite that a final transition is not triggered. Note
that conventionally TTs occur with gap closing,~\cite%
{Ying-2021-AQT,Topo-Wen,Hasan2010-RMP-topo,Yu2010ScienceHall,Chen2019GapClosing}
as those at level crossings. The TTs here around level anticrossings are
\textit{unconventional} ones in the sense that they occur without gap
closings,~\cite{Xie-QAH-2021,Amaricci-2015-no-gap-closing,Ying-gapped-top}
since both upper and lower gaps are finite at the transitions. The
unconventional TTs around level anticrossings become more popular in higher
excited states as illustrated by $j_E=10$ in Figure~\ref%
{fig-Nodes-Crossings-fix-Lambda}d-f where level crossings and aniticrossings
emerge more frequently.

\subsection{An Unconventional Topological Transition Universal for Different
Eigenstates}

\label{Sect-Transt-universal-jE}

When we vary the anisotropy, level crossings and anticrossings can also
occur similarly. Here, it is worthwhile to pay attention to a special TT, as
in \textbf{Figure}~\ref{fig-Nodes-Crossings-fix-g} around $\lambda=1.1 $,
which turns out to be quite particular with the following special features
together: (i) This transition is unconventional one without gap closing or
parity reversal, as demonstrated in Figure~\ref{fig-Nodes-Crossings-fix-g}%
a,b,d,e. (ii) Although this transition occurs in gapped situation, it seems
to be not associated with level anticrossing, as indicated by the dotted
line in Figure~\ref{fig-Nodes-Crossings-fix-g}f which is flat and shows no
sign of level anticrossing. (iii) While the variation of node number occurs
usually around the origin position, the new node for this transition is
coming from the infinity side, as shown Figure~\ref%
{fig-Nodes-Crossings-fix-g}a,d. (iv) This transition is universal for
different eigenstates, as illustrated for $j_E=1,5$ in Figure~\ref%
{fig-Nodes-Crossings-fix-g}a,d. This transition should originate from the
sign reversal of $g_y$ which changes the energy competitions for the Rashba
spin-orbit term without changing the parity.~\cite{Ying-gapped-top} Of
course, for this $g_y$-sign reversal to come to final effect it has to over
the resistance from the other terms in kinetic, potential, and tunneling
parts as in (\ref{Hx}). As the node in this transition is introduced from the infinity side, the state-dependent parity is not affected and
different states have similar wave-function profiles which are all decaying away from the main wave packets around the origin.
Thus, the $g_y$-sign reversal is facing an energy competition situation similar for different states, this may be the reason why this transition is universal.

This particular transition may have some special advantages. The gapped
situation in feature (i) avoids the detrimental slowing-down effect close to
a critical point with vanishing gap in preparing probe state if it is used
for some quantum sensors or devices.~\cite{Ying2022-Metrology} The feature
(ii) ensures a large gap, since otherwise level anticrossings have finite
but small gaps. The feature (iii) will not reverse the parity thus will not
introduce a level crossing or gap closing.~\cite{Ying-Stark-top} The feature
(iv) breaks the limitation of ground state so that eigenstates have the
same transition, which could provide great convenience in case the ground
state is difficult to reach. Moreover, the transition can occur in all
coupling regimes as one will see in Figure \ref{fig-Phase-Identity} and
Figure \ref{fig-phase-antiWinding}c.

\subsection{Phase Diagrams of Parity, Gaps and Nodes}

\label{Sect-Phase-diagrams}

\begin{figure}[t]
\includegraphics[width=1.0\columnwidth]{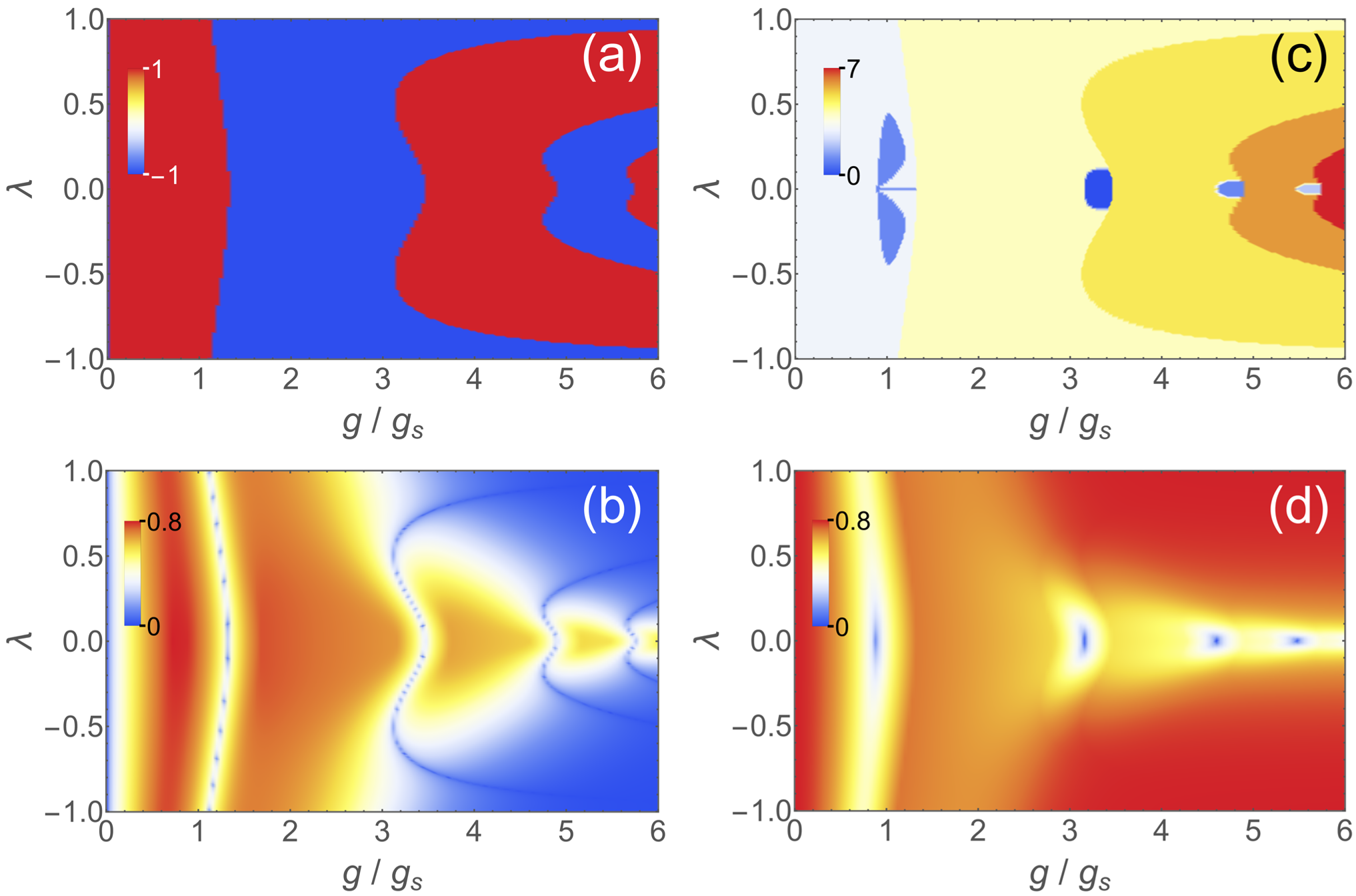}
\caption{Phase diagrams in $g$-$\protect\lambda$ plane for $j_E=5$. (a) Parity $P$, (b) Upper gap $\Delta_+/\Omega$%
, (c) Node number $n_Z$, (d) Lower gap $\Delta_-/\Omega$. }
\label{fig-PhaseDiag-jE=5}
\end{figure}
\begin{figure}[t]
\includegraphics[width=1.0\columnwidth]{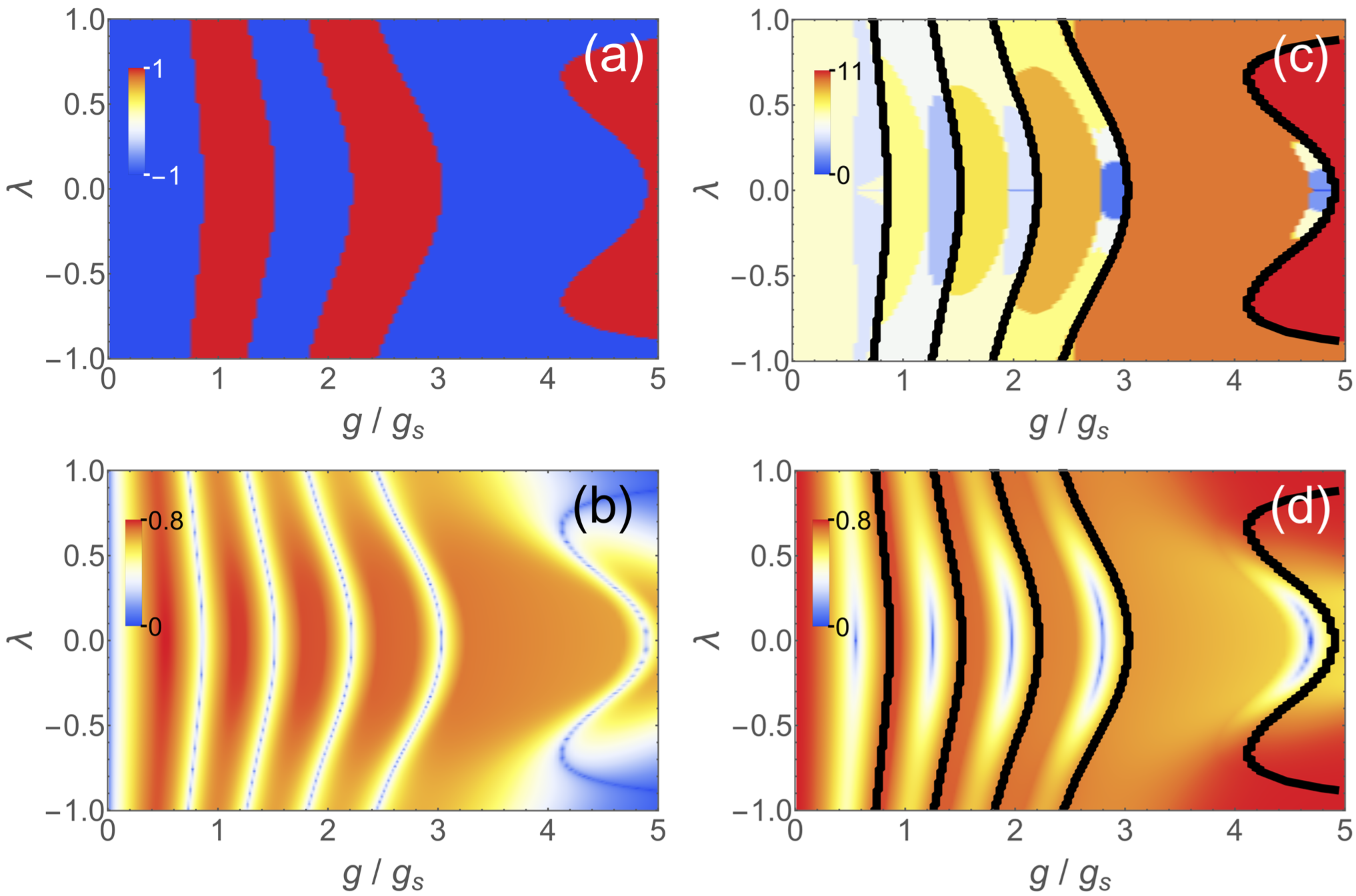}
\caption{Phase diagrams in $g$-$\protect\lambda$ plane for $j_E=11$. (a) Parity $P$, (b) Upper gap $\Delta_+/\Omega$%
, (c) Node number $n_Z$, (d) Lower gap $\Delta_-/\Omega$. The black solid
lines in (c,d) mark the level crossing boundaries in (a,b). }
\label{fig-PhaseDiag-jE=11}
\end{figure}

To have an overall view for the connection of the level
crossings/anticrossings and the TTs, we present the phase diagrams of
parity, gaps and nodes in $g$-$\lambda $ plane in \textbf{Figure}~\ref%
{fig-PhaseDiag-jE=5} for $j_{E}=5$ and \textbf{Figure}~\ref%
{fig-PhaseDiag-jE=11} for $j_{E}=11$. In Figure~\ref{fig-PhaseDiag-jE=5},
the parity reversal (panel (a)) occurs with upper gap closing (panel (b)) at
the main phase boundaries in the node phase diagram (panel (c)). The level
anticrossings take place apart from but nearby the level crossing
boundaries, as indicated by small lower gap $\Delta _{-}$ depicted by blue
color in panel (d). The anticrossing effect is weaker at larger couplings
around $g=3.1,4.6,5.5g_{\mathrm{s}}$, as one sees that the small-$\Delta _{-}
$ regions (blue) are narrow and only limited to the vicinity of $\lambda =0$%
, correspondingly the unconventional TTs with node-number variations in
Figure~\ref{fig-PhaseDiag-jE=5}c occur also around these narrow regions in
Figure~\ref{fig-PhaseDiag-jE=5}d. The level anticrossing effect is
relatively stronger around $g=0.9g_{\mathrm{s}}$ as the regime with the
small lower gap is more extended along $\lambda $ direction, while the
transitions in node number appear also in a wider regime. For a higher
excited state, as $j_{E}=11$ in Figure~\ref{fig-PhaseDiag-jE=11}, the
regimes of effective anticrossings are much extended for large couplings,
correspondingly the unconventional TTs merge in larger values of $\lambda $, as one
sees in panels (c) and (d).

The above analysis on the excited states also provides some more insight for
the ground state. The phase diagrams of parity, gaps and nodes for the
ground state have been established in refs.~\cite%
{Ying-2021-AQT,Ying-gapped-top}, as also will be re-classified in Section %
\ref{Sec-Full-Topo-Class-GS} of the present work after finding the
correspondence of nodes to the spin windings. What relevant here is the feature
that the TTs in the ground state are purely conventional ones except the
particular unconventional TT above $\lambda =1$ addressed in Section \ref%
{Sect-Transt-universal-jE}. The pure conventional TTs demonstrate that, as
far as the ground state is concerned, the anisotropic QRM and the JCM
topologically belong to the same class. This unification in some sense makes
up for the loss in the symmetry breaking of excitation number and
position-momentum duality~\cite{Ying-2021-AQT,Ying-Stark-top} which have
disconnected the JCM and the anisotropic QRM in the scaling relation in
critical universality classification.~\cite%
{LiuM2017PRL,Ying-2021-AQT,Ying-Stark-top} Now with the above analysis on
the level anticrossings, we gain a further understanding for the origin of
the pureness of topological phase transitions in ground states: there is no
anticrossing effect for the ground state as the only gap is the upper gap
which solely involves level crossings.

\begin{figure}[t]
\includegraphics[width=1.0\columnwidth]{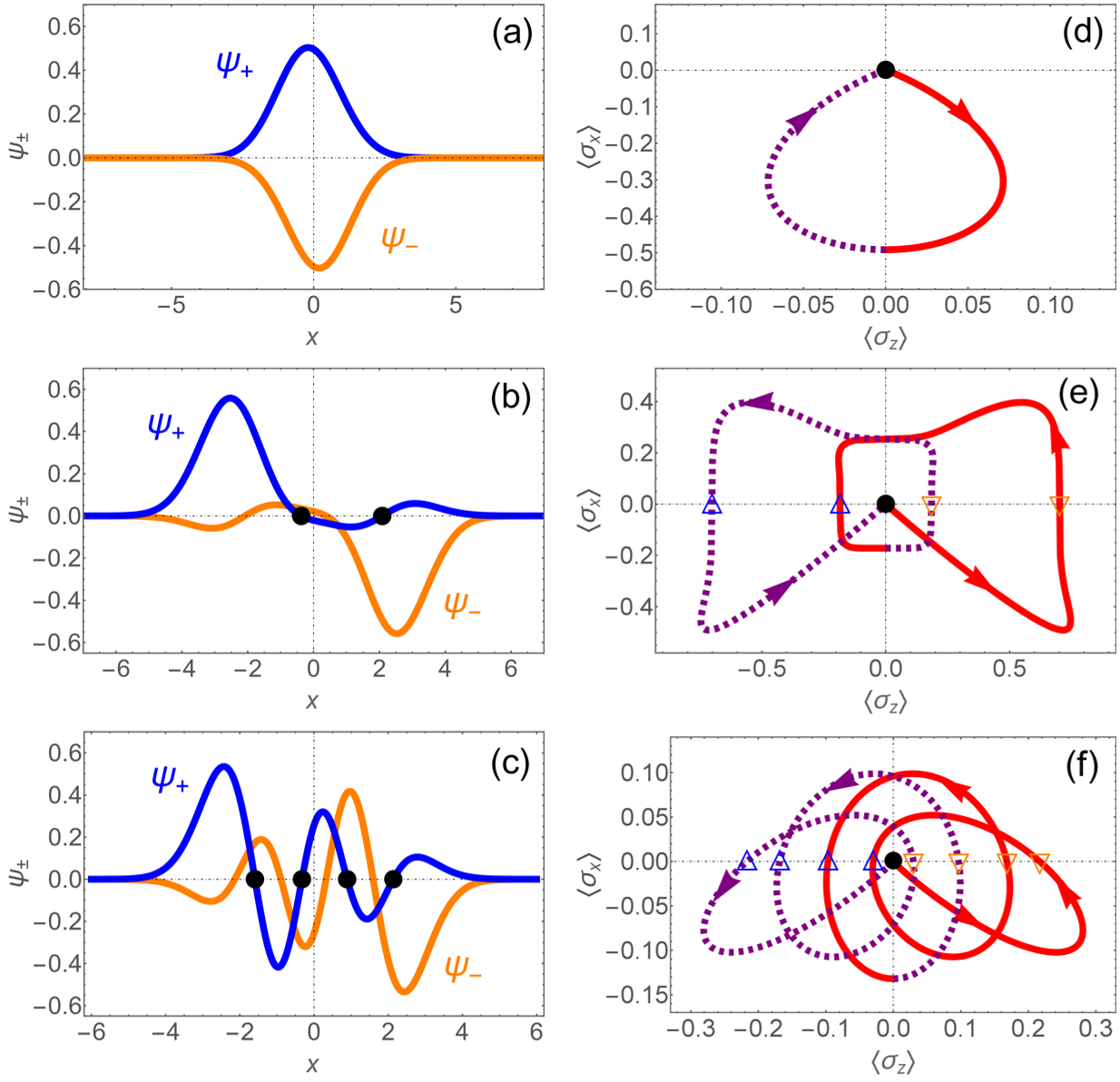}
\caption{\textit{Correspondence of nodes and spin windings in $0\leqslant
\protect\lambda<1$ regime:} Wave function $\psi _{\pm }\left( x\right)$ (a-c) and spin winding (d-f) of the ground state ($j_E=1$) at $%
g=1.5g_{\mathrm{s}}$ (a,d), the ground state at $g=4.4g_{\mathrm{s}}$ (b,e), and
the fifth state ($j_E=5$) at $g=1.5g_{\mathrm{s}}$ (c,f). The
dots in (a-c) mark the nodes in $\psi _{+}\left( x\right)$. The dots
in (d-f) locate the origin in $\langle \sigma_z\rangle$-$\langle
\sigma_x\rangle$ plane, the upward (downward) triangles label the nodes in
$\psi _{+}\left( x\right)$ ($\psi _{-}\left( x\right)$), and the arrows indicates the winding direction.
The $x<0$ ($x>0$) regime is plotted by solid (dotted) line in (d-f) and the
spin-expectation amplitudes are plotted by $| \protect\sigma_{i}|^{1/4}$ in
(e) for a better visibility. Here $\lambda=0.2$ for all panels. }
\label{fig-nodes-winding-1}
\end{figure}
\begin{figure}[t]
\includegraphics[width=1.0\columnwidth]{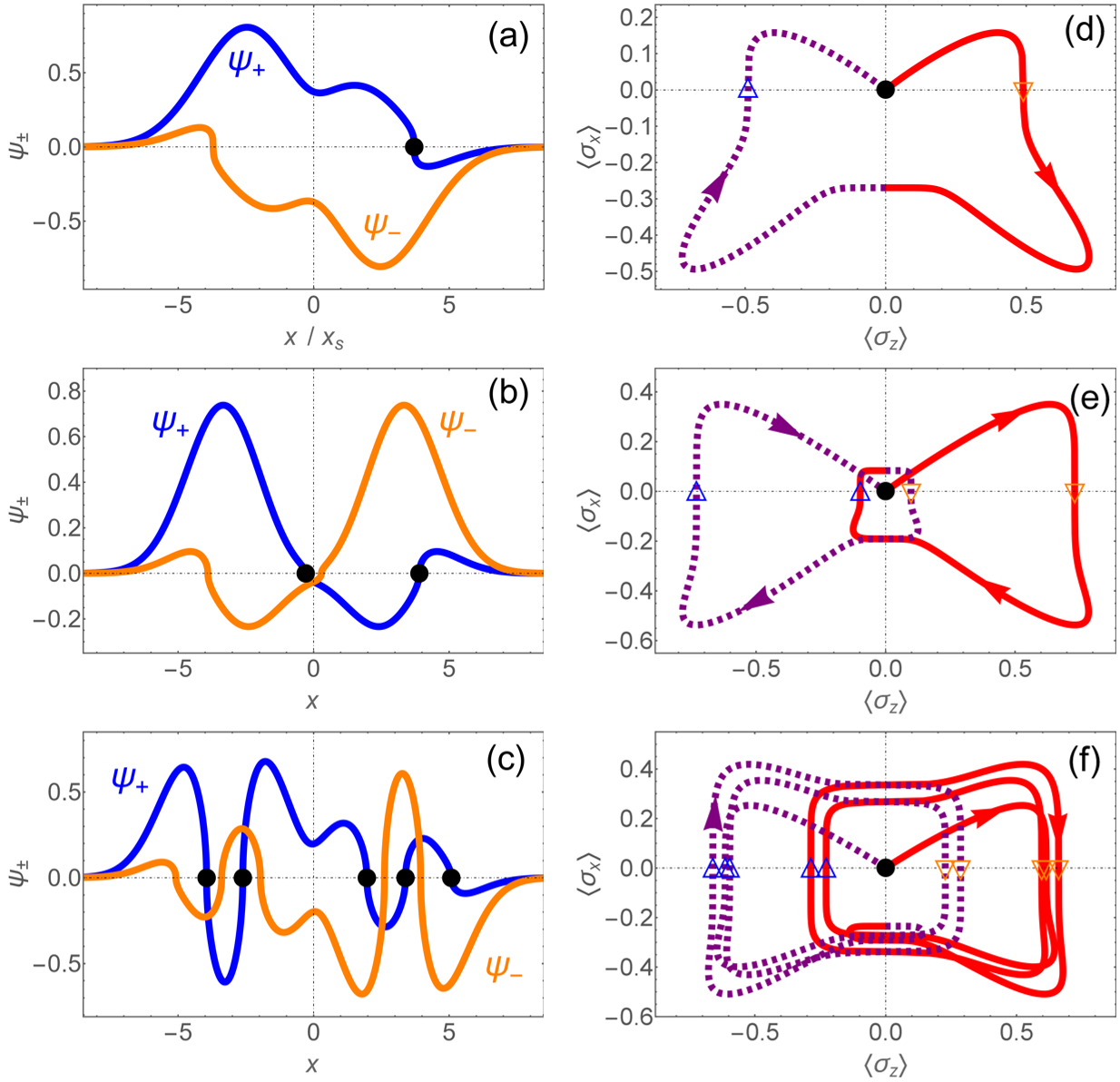}
\caption{\textit{Correspondence of nodes and spin windings in $\protect%
\lambda>1$ regime:} Wave function $\psi _{\pm }\left( x\right)$ (a-c) and spin winding (d-f) of
the ground state at $\protect\lambda=2.0$ (a,d), the ground state at $\protect\lambda=3.0$ (b,e), and the fifth
state at $\lambda=3.0$ (c,f), with the symbols as in Figure \ref{fig-nodes-winding-1}. The plot amplitudes are amplified by $| \protect\psi_{\pm}|^{1/3}$ in
(a), $| \psi_{\pm}|^{1/2}$ in (b,c), $| \sigma_{i}|^{1/4}$
in (d,f) and $| \sigma_{i}|^{1/5}$ in (e). Here $g=1.7g_{\mathrm{s}}$ for all panels. }
\label{fig-nodes-winding-2}
\end{figure}

\section{Spin Windings for Topological Feature}

\label{Sect-Spin-Windings}

The wave-function nodes might remind one of one-dimensional confined
spinless systems, the ground states of which however never have transition
of nodes due to the constraint of the no-node theorem.~\cite%
{Ref-No-node-theorem} Now that we also extend the discussion of TTs to
excited states, we should bring more attention to the fact that differently
from the spinless systems our systems involve the spin nontrivially,
especially the $g_y$ term resembles the Rashba/Dresselhaus spin-orbit
coupling while spin-orbit coupling is often fundamentally responsible for
TTs in condensed matter.~\cite%
{Hasan2010-RMP-topo,Yu2010ScienceHall,Nagasawa2013Rings,Ying2016Ellipse,Ying2017EllipseSC,Gentile2022NatElec}

In this section we shall look at the spin texture in the position space
which will upgrade the topological information from the wave function level
to physical observable level. The spin texture can be extracted by
\begin{eqnarray}
\langle \sigma _{z}\left( x\right) \rangle &=&\psi _{+}^{\ast }\left(
x\right) \psi _{+}\left( x\right) -\psi _{-}^{\ast }\left( x\right) \psi
_{-}\left( x\right) ,  \label{SpinZ-byWave} \\
\langle \sigma _{x}\left( x\right) \rangle &=&\psi _{+}^{\ast }\left(
x\right) \psi _{-}\left( x\right) +\psi _{-}^{\ast }\left( x\right) \psi
_{+}\left( x\right) ,  \label{SpinX-byWave} \\
\langle \sigma _{y}\left( x\right) \rangle &=&i\left[ \psi _{-}^{\ast
}\left( x\right) \psi _{+}\left( x\right) -\psi _{+}^{\ast }\left( x\right)
\psi _{-}\left( x\right) \right] ,  \label{SpinY-byWave}
\end{eqnarray}%
which are instantaneous spin expectations at position $x$. Note that, as
shown in Figure~\ref{fig-Ej}, the eigenstates are all non-degenerate except
at the crossing points, so that the eigenfunctions can be chosen to be
all real, i.e.
\begin{equation}
\psi _{\pm }^{\ast }\left( x\right) =\psi _{\pm }\left( x\right),
\label{WaveF-real}
\end{equation}
which is also valid even at the crossing points if one is not mixing up different
parity states. From the spin texture we will see the correspondence of
wave-function nodes and spin windings for the ground state and find more
hidden TTs in excited states.

\subsection{Constraints of the Parity Symmetry on Spin Texture}

\label{Sect-Parity-Constraints}

As afore-mentioned, the wave-function components in opposite spins are
related by $\psi _{-}\left( x\right) =P\psi _{+}\left( -x\right) $ due to
the parity symmetry. As a consequence, $\langle \sigma _{y}\rangle $ is
always vanishing,
\begin{equation}
\langle \sigma _{y}\left( x\right) \rangle =iP\left[ \psi _{+}\left(
-x\right) \psi _{+}\left( x\right) -\psi _{+}\left( x\right) \psi _{+}\left(
-x\right) \right] =0.
\end{equation}%
and the spin evolution lies only in $\langle \sigma _{z}\rangle $-$\langle
\sigma _{x}\rangle $ plane. Besides the vanishing $\langle \sigma _{y}\left(
x\right) \rangle ,$ the parity symmetry also leads to symmetric $\langle
\sigma _{x}\left( x\right) \rangle $ and antisymmetric $\langle \sigma
_{z}\left( x\right) \rangle $,
\begin{equation}
\langle \sigma _{x}\left( -x\right) \rangle =\langle \sigma _{x}\left(
x\right) \rangle ,\quad \langle \sigma _{z}\left( -x\right) \rangle
=-\langle \sigma _{z}\left( x\right) \rangle ,
\end{equation}%
which hold for both parity values $P=\pm 1$, as one can see directly from
Equations (\ref{SpinZ-byWave}) and (\ref{SpinX-byWave}).

\subsection{Correspondence of Wave-Function Nodes and Spin-Winding Zeros}

\label{Sect-Node-Winding-Corresp}

We find the evolution of $\langle \sigma _{z}\rangle $ and $\langle \sigma
_{x}\rangle $ with respect to $x$ forms spin windings in $\langle \sigma
_{z}\rangle $-$\langle \sigma _{x}\rangle $ plane, as shown in \textbf{Figure%
}~\ref{fig-nodes-winding-1} where the wave functions are shown in panels
(a-c) while the corresponding spin windings are displayed in panels (d-f).
At a small coupling where the wave function of the ground state has no node
in panel (a), also the spin winding does not effectively surround the origin
(black dot) in panel (d). At a larger coupling in panel (b) the wave
function has two nodes as marked by the black dots for $\psi _{+}$, the spin
in panel (e) is really winding around the origin with two pairs of zeros
of $\langle \sigma _{x}\rangle $ respectively corresponding to the
nodes in $\psi _{+}$ (upward triangles) and $\psi _{-}$ (downward
triangles). Similar spin windings can be seen for the excited states as
shown in panels (c,f) where more nodes are formed with more spin windings
correspondingly.

The correspondence of wave-function nodes and spin-winding zeros can be seen
from Equations (\ref{def-nodes}) and (\ref{SpinX-byWave}) which always
assign a vanishing value to $\langle \sigma _{x}(x)\rangle $
\begin{equation}
\langle \sigma _{x}\left( x_{Z}\right) \rangle =0
\end{equation}%
at a wave-function node $x_{Z}$ defined in (\ref{def-nodes}). Reversely a
zero of $\langle \sigma _{x}(x)\rangle $ also requires the node of wave
function as
\begin{equation}
\langle \sigma _{x}\left( x\right) \rangle =2\psi _{+}\left( x\right) \psi
_{-}\left( x\right)=2P\psi _{+}\left( x\right) \psi _{+}\left( -x\right).
\end{equation}
Any finite coupling will lead to a displacement in $v_{\sigma _{z}}(x)$ and
break the space-inversion symmetry (without spin reversal), thus in
principle one will not have negative-positive symmetric nodes within a same
component of wave function like $\psi _{+}(x_Z)=\psi _{+}(-x_Z)=0$ which
otherwise would have two nodes of $\psi _{+}(x)$ simultaneously correspond
to one $\langle \sigma _{x}(x)\rangle $ zero. In the absence of coupling one
has such negative-positive symmetric nodes, however even in such symmetric
cases $-x_z$ by parity symmetry is also the node of $\psi _{-}(x)$ which
avoids the double counting of the total node number in $\psi _{\pm}$. That
is to say, the total node number in $\psi _{\pm}$ is still equal to the
total zero number of $\langle \sigma _{x}(x)\rangle $. So we can conclude
that the wave-function nodes of $\psi _{\pm}$ and the $\langle \sigma
_{x}(x)\rangle $ zeros are one-to-one corresponding by total number or by
pair number. Thus, the wave-function nodes can be detected by the zeros of $%
\langle \sigma _{x}(x)\rangle $, which turns the topological information to
be measurable.

\subsection{Spin Winding Number}

\label{Sect-winding-number}

One can estimate the rounds of spin winding by the winding number around the
origin in the $\langle \sigma _{z}\rangle $-$\langle \sigma _{x}\rangle $
plane as calculated by
\begin{equation}
n_{zx}=\frac{1}{2\pi }\int_{-\infty }^{\infty }\frac{\langle \sigma
_{z}\left( x\right) \rangle \partial _{x}\langle \sigma _{x}\left( x\right)
\rangle -\langle \sigma _{x}\left( x\right) \rangle \partial _{x}\langle
\sigma _{z}\left( x\right) \rangle }{\langle \sigma _{z}\left( x\right)
\rangle ^{2}+\langle \sigma _{x}\left( x\right) \rangle ^{2}}dx,
\label{n-zx}
\end{equation}%
which has also been applied in topological classification in nanowire
systems and quantum systems with geometric driving.~\cite%
{Ying2016Ellipse,Ying2017EllipseSC,Ying2020PRR,Gentile2022NatElec} It should
be reminded here that the wave function for $\lambda <0$ regime is defined
in momentum space as mentioned in Sect.~\ref{Sect-Model}, thus we should
also take spin texture in momentum space where the spin winding is in
$\langle \sigma _{y}\rangle $-$\langle \sigma _{x}\rangle $ plane.

Here by Equation (\ref{n-zx}) a problem arises from the boundary condition
as there is a fractional winding angle at infinity. Nevertheless, we can
neglect this boundary angle and simply take the nearest integer of $n_{zx}$
as a final effective winding number
\begin{equation}
n_{w}=\text{nint}(n_{zx}),  \label{n-w}
\end{equation}%
where the rounding function $\text{nint}(x)$ gives the integer closest to $x$%
. This is equivalent to approximately regarding the
spin winding in $\langle \sigma_z\rangle$-$\langle
\sigma_x\rangle$ plane as connected at the two ends of $x \rightarrow \pm\infty$ as if periodic condition. In this way, the effective spin winding number $n_{w}$ is equal to the
wave-function node number $n_{Z}$ in the topological phases of the ground
state.

\begin{figure}[t]
\includegraphics[width=0.9\columnwidth]{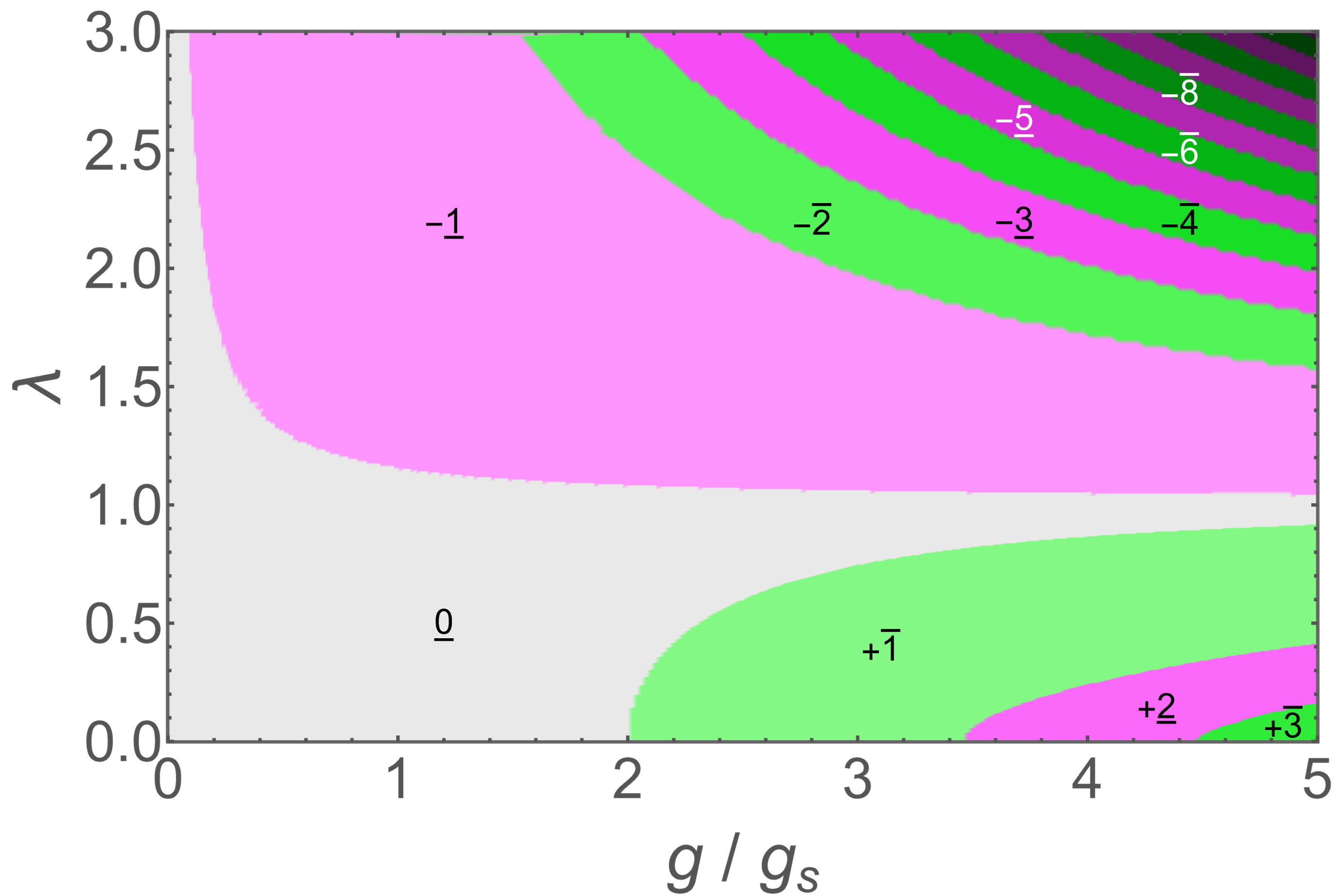}
\caption{\textit{Spin winding direction and a full classification in
topological phase diagram of ground state.} The number represents the node
number or the spin winding number with the positive (negative) sign labeling
counterclockwise (clockwise) winding direction, while the overline (underline)
denotes positive (negative) parity.}
\label{fig-Phase-Identity}
\end{figure}

\subsection{Spin Winding Directions}

\label{Sect-winding-direction}

Besides the rounds of the spin winding, the winding direction turns out to
be another important feature. The examples shown in Figure~\ref%
{fig-nodes-winding-1} are illustrated in $0<\lambda <1$ regime, where the
effective winding direction for the low-lying states with a finite number of
nodes is more probably counterclockwise like the direction of the polar
angle, as indicated by the arrows along the spin trajectories in Figure~\ref%
{fig-nodes-winding-1}e,f. The effective winding direction can also be
clockwise, which happens more probably in $\lambda >1$ regime as the
effective Rashha field $g_{y}$ changes the sign beyond $\lambda =1$. We show
some examples in \textbf{Figure}~\ref{fig-nodes-winding-2} both for the
ground state and for excited states. Here, besides the similar
correspondence of nodes and spin winding zeros, one sees that the spin is indeed
winding in clockwise direction. Now we realize that the
clockwise/counterclockwise winding direction is also a quantum feature that
should be emphasized in topological identifications of quantum states in
light-matter interactions. Of course, the winding direction can be reflected
in the sign of $n_{w}$ if one does not take the winding number only by its
amplitude.

\subsection{A Full Topological Classification for Ground State}
\label{Sec-Full-Topo-Class-GS}

Inspired by the above analysis on the nodes and the spin windings, we can
combine the parity, the node number and effective winding number to identify
different quantum states of the ground state. We present the ground-state
phase diagram with a full topological identification in \textbf{Figure}~\ref%
{fig-Phase-Identity}. Here the numbers represent the node number $n_{Z}$,
which is equal to the absolute effective spin-winding number $n_{w}$ for the
ground state, the overline (underline) denotes positive (negative) parity,
while the positive (negative) sign before the numbers denotes
counterclockwise (clockwise) spin winding direction. One can recognize the
counterclockwise (clockwise) spin windings in $\lambda <1$ ($\lambda <1$)
regime. The special unconventional TT mentioned in Sect \ref%
{Sect-Transt-universal-jE} occurs at the $\underline{0}$/$\left( -\underline{%
1}\right) $ boundary where there is neither gap closing nor parity reversal,
in contrast to the other boundaries which are conventional ones.

\section{Big (Bridge) Spin Knots for Novel Topological Transitions}

\label{Sect-Bridge-Knot-TopTrans}

\subsection{Anti-Winding Nodes from Big (Bridge) Spin Knots in Excited States%
}

\begin{figure}[t]
\includegraphics[width=1.0\columnwidth]{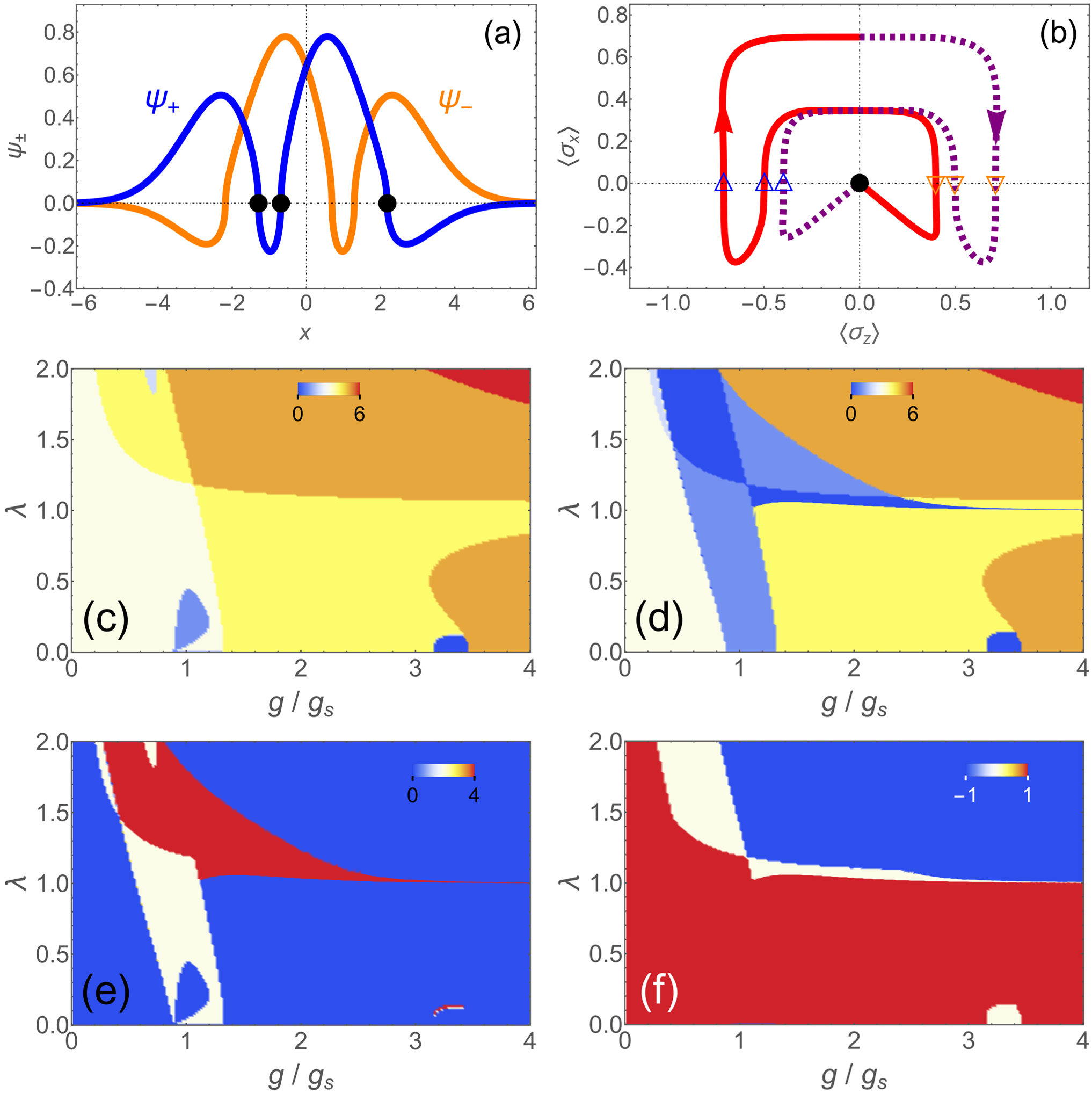}
\caption{\textit{Anti-winding-node transitions in excited states.}
a,b) Nodes (dots) of the wave function (a) and zeros (triangles) of spin winding in $\langle
\sigma_z\rangle$-$\langle \sigma_x\rangle$ plane (b) at $\lambda =0.8, g=0.9g_{\rm s}$.
c-f) Phase diagrams of $n_Z$ (c), $|n_w|$ (d), $n_Z-|n_w|$ (e) and $\text{sign}(n_w)$ (f) in $g$-$\lambda$ plane.
Here $j_E=5$ and the plot amplitude is amplified by $|\psi _{\pm}(x)|^{1/2}$ and $\langle \sigma_{x,z} (x)\rangle^{1/3}$ in (a,b).}
\label{fig-phase-antiWinding}
\end{figure}

In last section we have seen the correspondence not only of the
wave-function nodes and the spin-winding zeros but also of the node number $%
n_{Z}$ and the spin winding number $n_{w}$ in the ground state. In excited
states, the nodes-zeros correspondence from the wave functions and the spin
windings always holds but the $n_{Z}$-$n_{w}$ correspondence may be broken,
which unveils a different kinds of TTs. Indeed, besides the spin winding
around the origin, the spin trajectory can also form spin knots that are
not surrounding the origin. Note such spin knots have no contribution to the
spin winding number as the spin trajectory in a spin knot comes back to a
same point which cancels the winding angle. When the spin knot is small and
not crossing the $\langle \sigma _{z}\rangle $ axis, as the case in Figure %
\ref{fig-nodes-winding-2}f where actually there is a small knot below the origin, the
winding number is not affected and $n_{Z}$ is equal to $n_{w}$. However, the
spin knot can be also large so that the spin trajectory can cross the $%
\langle \sigma _{z}\rangle $ axis to form spin-winding zeros, as in \textbf{%
Figure}~\ref{fig-phase-antiWinding}b where the knot forms a bridge-like
profile with two piers crossing the $\langle \sigma _{z}\rangle $ axis. In
this situation the final spin winding number is smaller than the
wave-function node number as in Figure \ref{fig-phase-antiWinding}a. We can
name such a knot by \textit{big spin knot} or \textit{bridge spin knot} and
the involved nodes\ by \textit{anti-winding nodes} which do not contribute
to the spin windings.

\subsection{Anti-Winding-Node Topological Transitions}

When a bridge knot with anti-winding nodes is formed, a new topological
transition occurs concerning the difference of the wave-function node number
and the spin winding number. Figure \ref{fig-phase-antiWinding}c shows the
phase diagram of $n_{Z}$ for $j_{E}=5$ at $\omega =0.5\Omega $, with the
anisotropy strength extended to $\lambda =2$. Here the vertically tilting
line around $g=1.0\sim 1.3g_{\mathrm{s}}$ is the first level crossing line
in Figure \ref{fig-PhaseDiag-jE=5}. The corresponding map of $n_{w}$ is
presented in Figure \ref{fig-phase-antiWinding}d in which three new main
boundaries appear: (i) around $g=0.3\sim 0.9g_{\mathrm{s}}$, vertically
parallel to the first level crossing line; (ii) horizontally around $\lambda
\sim 1$; (iii) the left slash line in $\lambda >1$ and $g\gtrsim 1.0g_{%
\mathrm{s}}$ regime. Note the node number $n_{Z}$ remains unchanged across
all these three boundaries while the winding number $n_{w}$ has a jump,
which indicates a new type of TTs. The transition at boundary (i) should be
connected with the level aniticrossing as the boundary is around the level
aniticrossing in Figure \ref{fig-PhaseDiag-jE=5}d, thus the level
aniticrossing is associated not only with node topological transitions but
also with anti-winding-node TTs. The transition at boundary (ii) should
originate from the sign reversal of $g_{y}$ around $\lambda =1$ which
changes the relation of the Rashba spin-orbit term and tunneling energy from
counteracting to competing.\cite{Ying-2021-AQT} The transition at boundary
(iii) may come from a larger $g_{y}$ that renders the Rashba spin-orbit
coupling to dominate over the tunneling.

The discount of winding number $n_{w}$ by the \textit{anti-winding} spin knots can be
seen in Figure \ref{fig-phase-antiWinding}e which shows
\begin{equation}
n_{aw}=  n_{Z}  -\left\vert n_{w}\right\vert ,
\label{n-AntiW-knots}
\end{equation}%
denoting the pair number of nodes involved in bridge spin knots. In the blue
region $ n_{Z} $ and $\left\vert n_{w}\right\vert $ are
equal and there is no bridge spin knot, while there is a discount of $%
\left\vert n_{Z}\right\vert $ in other regions where bridge spin knots
appear.

The total winding direction is indicated by the sign of $n_{w}$ in Figure %
\ref{fig-phase-antiWinding}f. We see here that the spin is winding
counterclockwise in most $\lambda <1$ regime and also in $\lambda >1$ regime
with small $g$, while clockwise spin winding is found for large $g$ in $%
\lambda >1$ regime. In the white regions, the spin effectively does not wind around the origin.

\subsection{Topological Multiple Points in Excited States}

At this point it may be worthwhile to mention the topological multiple points where
different topological boundaries meet. The ground state of the anisotropic QRM has a
multicritical point and series of quadruple points along $\lambda =0$
line which however are not topological
multiple points but crossings of transitions in topological class and Landau class.~\cite{Ying-2021-AQT,Ying-gapped-top}
While there is a symmetry breaking in traditional transitions of Landau class, TTs preserve the symmetry.
Indeed, as one sees in Figure~\ref{fig-Phase-Identity},
the topological boundaries do not cross each other in the ground state,
although the presence of the nonlinear Stark coupling can create topological
quadruple points.~\cite{Ying-Stark-top} Here from Figures
\ref{fig-PhaseDiag-jE=5}c,\ref{fig-PhaseDiag-jE=11}c,\ref{fig-phase-antiWinding}c-e
we see that the topological multiple points can occur in the excited
states without introducing the Stark coupling. These topological multiple points have various crossings among
unconventional TT boundaries, conventional ones, and
anti-winding-node ones. The regimes around the topological multiple points are
topologically sensitive to the parameter variations which might be useful to
design topological quantum devices or sensors, while different types of
multiple points could provide more varieties or choices in need.

\section{Node Sorting and Algebraic Formulation for General Spin Winding
Number with Spin Knots}

\label{Sect-Node-Sort-Algebraic-Formul}

\begin{figure*}[t]
\includegraphics[width=2.0\columnwidth]{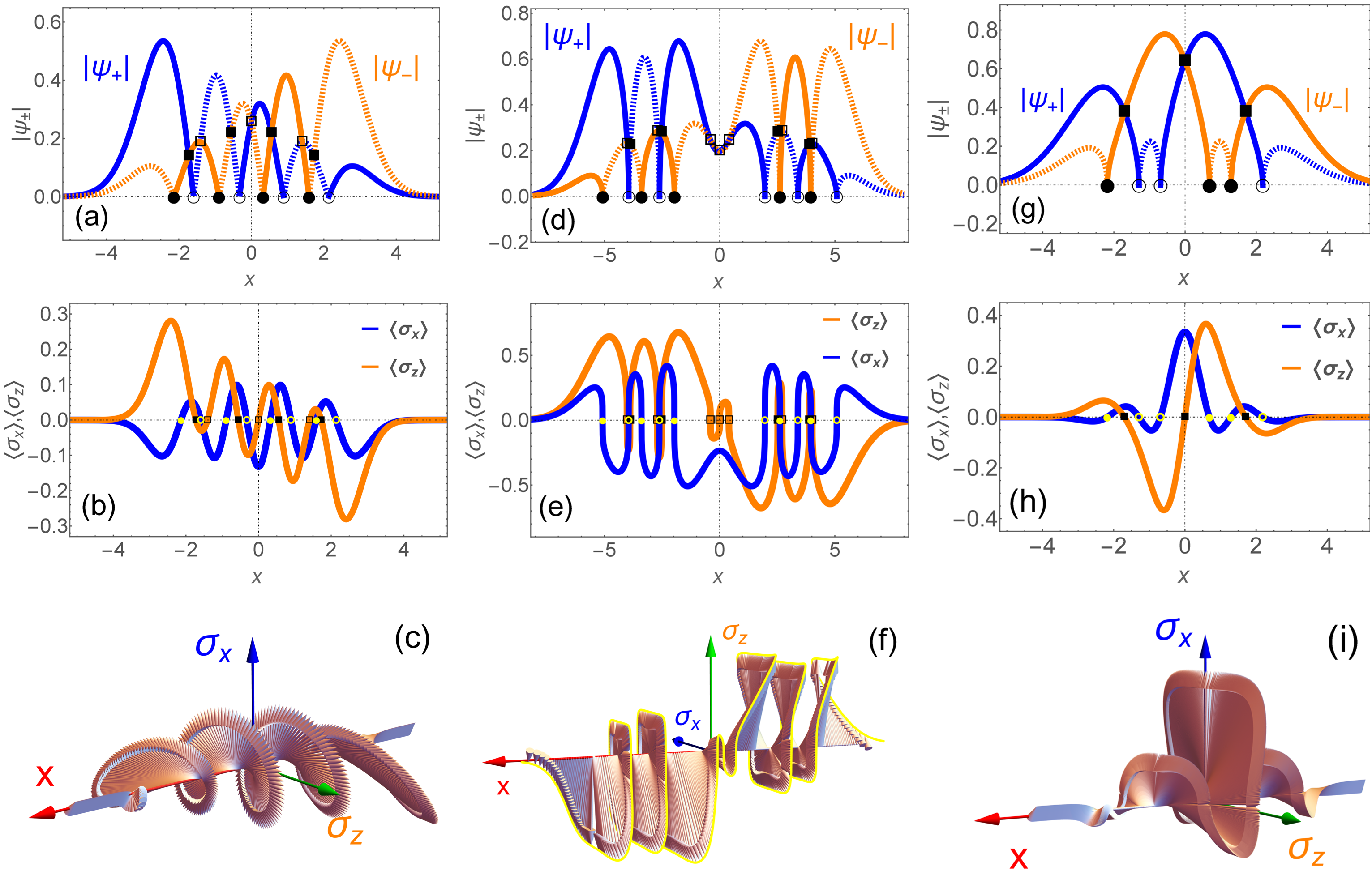}
\caption{\textit{Node distribution order and algebraic formulation for
general spin windings.} Wave-function amplitudes $| \protect\psi _{+}(x)|$
(blue) and $| \protect\psi _{-}(x)|$ (orange) (a,d,g), instantaneous spin
expectations $\langle \protect\sigma _{x}\left( x\right) \rangle $ (blue)
and $\langle \protect\sigma _{z}\left( x\right) \rangle $ (orange) (b,e,h),
three-dimensional spin texture (c,f,i) for $j_E=5$ at: a-c) $\protect\lambda =0.2$ and $g=1.5g_{\mathrm{s}}$
without spin knot, corresponding to Figure \ref{fig-nodes-winding-1}f; d-f) $\protect\lambda =3.0$ and $g=1.7g_{\mathrm{s}}$
with a small knot, corresponding to Figure \ref{fig-nodes-winding-2}f; g-i) $\protect\lambda =0.8$ and $g=0.9g_{\mathrm{s}}$
with a big knot and anti-winding nodes, corresponding to Figure \ref{fig-phase-antiWinding}b. In (a,d,g) solid (dotted) lines represent $%
\protect\psi _{\pm}(x)=| \protect\psi _{\pm}(x)|$ ($\protect\psi %
_{\pm}(x)=-| \protect\psi _{\pm}(x)|$). In (a,d,g) circles (squares) are
nodes of $\protect\psi _{\pm }$ ($\widetilde{\protect\psi} _{\pm }$),
correspondingly in (b,e,h) filled (empty) circles mark $\langle \sigma _{x}(x) \rangle $ zeros with positive (negative) $\langle
\sigma _{z}(x) \rangle $, while filled (empty) squares denote $\langle \sigma _{z}(x) \rangle $ zeros with positive (negative) $\langle \sigma _{x}(x) \rangle $. }
\label{fig-nW-by-sZsX}
\end{figure*}

\subsection{Another Set of Wave-Function Nodes on Spin-$\protect\sigma _{x}$
Basis}

So far we have only considered the nodes in the wave function component $%
\psi _{\pm }\left( x\right) $ on spin-$\sigma _{z}$ basis $\left\vert
\uparrow \right\rangle ,\left\vert \downarrow \right\rangle $. We can also
transform to spin-$\sigma _{x}$ basis $\left\vert \Uparrow \right\rangle
=\left( \left\vert \uparrow \right\rangle +\left\vert \downarrow
\right\rangle \right) /\sqrt{2}$, $\left\vert \Downarrow \right\rangle
=\left( \left\vert \uparrow \right\rangle -\left\vert \downarrow
\right\rangle \right) /\sqrt{2}$ by $\left\vert \psi \left( x\right)
\right\rangle =\psi _{+}\left( x\right) \left\vert \uparrow \right\rangle
+\psi _{-}\left( x\right) \left\vert \downarrow \right\rangle =\widetilde{%
\psi }_{+}\left( x\right) \left\vert \Uparrow \right\rangle +\widetilde{\psi
}_{-}\left( x\right) \left\vert \Downarrow \right\rangle $ where%
\begin{equation}
\widetilde{\psi }_{\pm }\left( x\right) =\frac{1}{\sqrt{2}}\left[ \psi
_{+}\left( x\right) \pm \psi _{-}\left( x\right) \right] .
\end{equation}%
Note $\widetilde{\psi }_{\pm }\left( x\right) $ can also have nodes $%
x=y_{Z,j}$ which actually correspond to the zeros of $\langle \sigma
_{z}\left( x\right) \rangle $%
\begin{equation}
\widetilde{\psi }_{+}\left( y_{Z,j}\right) \text{ or }\widetilde{\psi }%
_{-}\left( y_{Z,j}\right) =0,\quad \langle \sigma _{z}\left( y_{Z,j}\right)
\rangle =0,
\end{equation}%
at equal wave-function amplitudes of $\psi _{\pm }\left( x\right) $,
\begin{equation}
\left\vert \psi _{+}\left( y_{Z,j}\right) \right\vert =\left\vert \psi
_{-}\left( y_{Z,j}\right) \right\vert ,  \label{Eq-Amplitude}
\end{equation}%
as concluded from Equations (\ref{SpinZ-byWave}) and (\ref{WaveF-real}).
Note here under the parity symmetry the two wave-function components $%
\widetilde{\psi }_{\pm }\left( x\right) =\frac{1}{\sqrt{2}}\left[ \psi
_{+}\left( x\right) \pm P\psi _{+}\left( -x\right) \right] $ have different
space inversion symmetries (symmetric and anti-symmetric), thus having
different node situations. It is more convenient to take (\ref{Eq-Amplitude}%
) for locations of $y_{Z,j}$. As addressed below we find the topological
information of the spin windings and spin knots is encoded in the order
of the $\psi _{\pm }\left( x\right) $ nodes $x_{Z,i}$ and the $\widetilde{%
\psi }_{\pm }\left( x\right) $ nodes $y_{Z,j}$, where $i,j$ are number
labeling for multi-node case.

\subsection{Node Sorting and Algebraic Formulation of Spin Winding Number}
\label{sec-nw-Algeb-Formu}

Since the spin winding will finally go through these nodes or zeros, we propose an
alternative way to calculate the winding number in terms of the node numbers
of $\psi _{\pm }\left( x\right) $ and $\widetilde{\psi }_{\pm }\left(
x\right) $ or equivalently the zero numbers of $\langle \sigma _{x}\left(
x\right) \rangle $ and $\langle \sigma _{z}\left( x\right) \rangle $.

Suppose $x_{Z,1}<x_{Z,2}\cdots <x_{Z,2n_{Z}}$ are the nodes of $\psi _{\pm
}\left( x\right) $ or equivalently the zeros of $\langle \sigma _{x}\left(
x\right) \rangle $, then we introduce $n_{Z}^{(+,-)}$ to count the nodes
among the original ones $\{x_{Z,k}|\ k=1,\cdots , 2n_{Z}\}$ that change the
sign of $\langle \sigma _{z}(x_{Z,k}\rangle $, i.e. $%
x_{Z,1}=x_{Z,1}^{(+,-)}<x_{Z,2}^{(+,-)}\cdots <x_{Z,n_{Z}^{(+,-)}}^{(+,-)}$
at which $\langle \sigma _{z}(x_{Z,i}^{(+,-)})\rangle \times \langle \sigma
_{z}(x_{Z,i+1}^{(+,-)})\rangle <0$. We denote the sign of $\langle \sigma _{z}(x_{Z,i}^{(+,-)})\rangle $ by  $S^{(+,-)}_{i}$.

Let $m(i)$ count the nodes $y_{Z,j}$ ($%
j=1,\cdots ,m(i)$) of $\langle \sigma _{z}\left( x\right) \rangle $ (not $\langle \sigma _{x}\left( x\right) \rangle $) in the
section $[x_{Z,i}^{(+,-)},x_{Z,i+1}^{(+,-)}]$ and each node $y_{Z,j}$ acquires a negative
(positive) sign denoted by $s^{i}_j$ for negative (positive) $\langle \sigma _{x}\left(
y_{Z,j}\right) \rangle $. Generally $m(i)$
is an odd number in an effective contribution to the final winding number,
since an even $m(i)$ means the spin trajectory is returning thus cancels the
winding angle.

With the above node sorting we obtain an algebraic formula for the spin
winding number
\begin{eqnarray}
n_{w} &=&n_{w}^{Z}\equiv \sum_{i=1}^{n_{Z}^{(+,-)}}n_{w,i}^{Z},
\label{nW-by-z} \\
n_{w,i}^{Z} &=&\frac{1}{4}\left[ 1-\left( -1\right) ^{m(i)}\right] S^{(+,-)}_{i} s^{i}_{m(i)} ,
\end{eqnarray}%
as a replacement of the integral form of spin winding number in Equations (%
\ref{n-zx}) and (\ref{n-w}). Here, as mentioned in Section \ref{Sect-winding-number}, we regard the spin winding as connected at the two infinity ends $x\rightarrow\pm \infty$ in $\langle \sigma_z\rangle$-$\langle
\sigma_x\rangle$ plane, thus the $n_{Z}^{(+,-)}$'th section is composed of $[x_{Z,n_{Z}^{(+,-)}}^{(+,-)},\infty)$ and $(-\infty,x_{Z,1}]$. Then the two infinity ends together contribute once to $m(n_{Z}^{(+,-)})$ with the sign equal to that of $\langle \sigma _{x}(x)\rangle |_{x\rightarrow \infty }$. Of course, in principle $m(n_{Z}^{(+,-)})$ can also be larger than $1$, since in the $n_{Z}^{(+,-)}$'th section there could be $\langle \sigma _{z}\left( x\right) \rangle $ zeros besides the contribution of the connected infinity boundary.

\subsection{Illustrations of Node Sorting in the Absence/Presence of Spin
Knots}

To see more clearly the relation of spin winding number and the node
sorting, in \textbf{Figure} \ref{fig-nW-by-sZsX} we illustrate some typical
examples of wave-function amplitudes $\left\vert \psi _{\pm }(x)\right\vert $
and spin textures, without spin knots (a-c), with a small knot (d-f), and
with a big (bridge) knot with anti-winding nodes (g-i), respectively
corresponding to Figures \ref{fig-nodes-winding-1}f, \ref%
{fig-nodes-winding-2}f, and \ref{fig-phase-antiWinding}b.

In Figure \ref{fig-nW-by-sZsX}a,d,g the nodes of $\psi _{\pm }(x)$ are
marked by black filled (empty) circles which correspond to the zeros of $%
\langle \sigma _{x}(x)\rangle $ while $\langle \sigma _{z}(x)\rangle $ is
positive (negative), as determined by the relations $\langle \sigma
_{x}\left( x\right) \rangle =2\psi _{+}\left( x\right) \psi _{-}\left(
x\right) $ and $\langle \sigma _{z}\left( x\right) \rangle =\left\vert \psi
_{+}\left( x\right) \right\vert ^{2}-\left\vert \psi _{-}\left( x\right)
\right\vert ^{2}$. We see that all $\psi _{+}\left( x\right) $ nodes (empty
circles) have negative $\langle \sigma _{z}\left( x\right) \rangle $ while
all $\psi _{-}\left( x\right) $ nodes (filled circles) have positive $%
\langle \sigma _{z}\left( x\right) \rangle$. On the other hand, the nodes of
$\widetilde{\psi }_{\pm }\left( x\right) $ or the zeros of $\langle \sigma
_{z}\left( x\right) \rangle $ are located at points where $\left\vert \psi
_{+}\left( x\right) \right\vert $ and $\left\vert \psi _{-}\left( x\right)
\right\vert $ are equal, either with same signs (filled squares) of $\psi
_{+}\left( x\right) $ and $\psi _{-}\left( x\right) $ at crossings of both
solid lines or both dotted lines or with opposite signs (empty squares) of $%
\psi _{+}\left( x\right) $ and $\psi _{-}\left( x\right) $ at crossings of
solid lines and dotted lines, the former yields positive $\langle \sigma
_{x}\left( x\right) \rangle $ while the latter gives negative $%
\langle \sigma _{x}\left( x\right) \rangle $. The filled/empty
circles and squares in Figure \ref{fig-nW-by-sZsX}b,e,h represent the zeros
and signs of $\langle \sigma _{x}\left( x\right) \rangle $ and $\langle
\sigma _{z}\left( x\right) \rangle $ the same as in Figure \ref%
{fig-nW-by-sZsX}a,d,g.

Figure \ref{fig-nW-by-sZsX}a,b show a completely alternate node or zero
sequence,
filled-circle/filled-square/empty-circle/empty-square/filled-circle/...,
indicating full spin windings without any spin knot as in two-dimensional
Figure~\ref{fig-nodes-winding-1}f and three-dimensional Figure~\ref%
{fig-nW-by-sZsX}c. Figure \ref{fig-nW-by-sZsX}d-f include a small spin knot
(in $\langle \sigma _{z}(x)\rangle $-$\langle \sigma _{x}(x)\rangle $ plane
in Figure \ref{fig-nodes-winding-2}f) around $x=0$ as displayed by the
successive empty squares (zeros of $\langle \sigma _{z}\left( x\right)
\rangle $) during the turns of spin trajectory. Nevertheless, the successive
squares appear between a same pair of filled and empty circles, thus not
affecting the equality of the winding number $n_{w}$ and the node number $%
n_{Z}$. However, the successive filled squares in Figure \ref%
{fig-nodes-winding-1}g,h show up in different pairs of filled and empty
circles, which invalidates the node number in contribution to the winding
number. Finally the two cases in Figure \ref{fig-nW-by-sZsX}a-f have
$\left\vert n_{w}\right\vert =\left\vert n_{w}^{Z}\right\vert=n_{Z}$ while the case in Figure \ref{fig-nW-by-sZsX}g-i has
$\left\vert n_{w}\right\vert =\left\vert n_{w}^{Z}\right\vert <\left\vert
n_{Z}\right\vert $.

\subsection{Topological Identity and Bridge of Geometric Topology and Algebraic Topology}
\label{sec-Topo-Id}

One may find the convenience of Equation (\ref{nW-by-z}) that by replacement
of $n_{w}$ with $n_{w}^{Z}$ it numerically avoids the more difficult
integral in the original winding expression (\ref{n-zx}) for $n_{w}$ which involves derivatives. More
importantly it demonstrates the relation of wave-function nodes and spin
windings in an explicit and solid way. The integral in $n_{w}$ depends on
the topological structure of spin trajectory geometrically, while the nodes
and the sorting involved in $n_{w}^{Z}$ reflect the topological information
of the wave function or the spin texture algebraically. The equation also
reclaims the original sense of topological classification that no matter how
the spin trajectory is geometrically deformed one has the same topological
winding number as long as the few points of node sequence are given, just as
one cannot change a torus into a sphere by a continuous deformation in the
well-known illustration for topological difference where the holes of an
object are given.

Besides the algebraic nature of the nodes themselves as in a function, one
can also relabel the nodes with four numbers so that the node topological
information is more clearly represented by the a sequence code of the four
numbers. Different sorting and ordering of the numbers give different
topology, which is a kind of algebraic topology. For an example, we can
assign the numbers as these following the symbols: filled circle ($1$),
filled square ($2$), empty circle ($3$) and empty square ($4$). Then the
node number code for a smooth counterclockwise winding as in Figure \ref%
{fig-nW-by-sZsX}a-c is $123412341234123\widetilde{4} $, while exchange of $2$ and $4$
gives a clockwise winding $\cdots 1432 1432 \cdots $. Here, the digit with tilde is coming from the two infinity ends which are assumed to be connected in Section \ref{sec-nw-Algeb-Formu} and $\widetilde{2}$ ($\widetilde{4}$) stands for same (opposite) sign of $\psi_+(x)$ and $\psi_-(x)$ or equivalently positive (negative) value of $\langle \sigma _{x}(x)\rangle$ at $x\rightarrow \pm\infty$. Each period means a
round of winding. A section with adjacently repeating 2 or 4 denotes a small
knot as in Figure \ref{fig-nW-by-sZsX}d-f which have a clockwise code $%
1432 1432 144432 1432 143\widetilde{2} $ with three adjacent $4$s in the middle. An
anti-winding big spin knot will break the code order either $1234$ or $1432$%
, as in Figure \ref{fig-nW-by-sZsX}g-i which are embedded a disordered code $%
123321123\widetilde{4}$. Such a node code provides a topological identity to label the topological structure of the eigenstate.

In such a sense, Equation (\ref{nW-by-z}) also builds the bridge of
geometric topology and algebraic topology, but in a physical way as the
nodes are originally of the wave function and the windings are of the spin
in a quantum mechanical system. The mapping of topological structure of the
quantum states onto number sequences may provide a way for topological
quantum encoding and decoding.

\begin{figure*}[t]
\includegraphics[width=2.0\columnwidth]{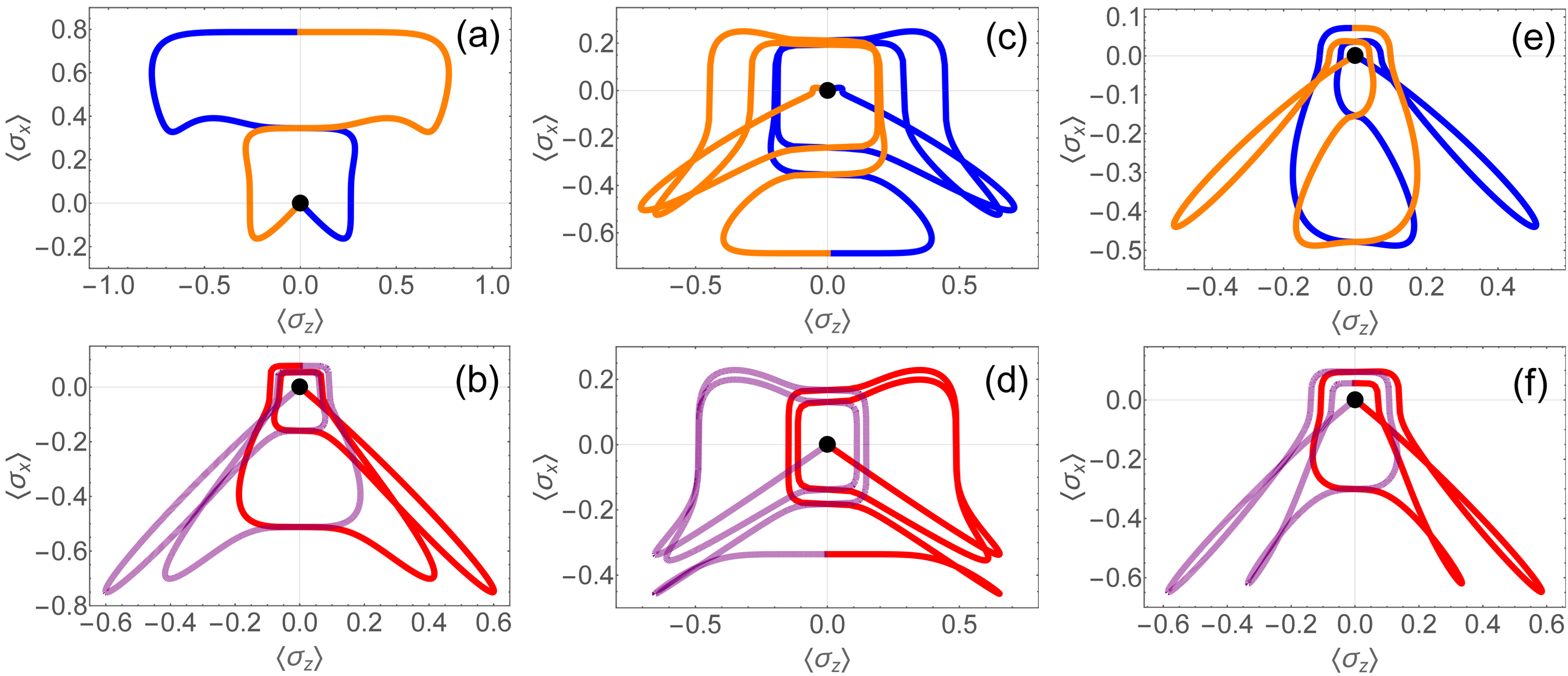}
\caption{\textit{Typical spin windings with different kinds of knots in excited states:
Anisotropic QRM as a born topological abstract artist.} Spin windings in
formal or spiritual similarity with a) a chef in a biggest hat, and with
beard ($g=1.0g_s, \protect\lambda=0.2$); b) a sumo wrestler ($g=0.1g_s,
\protect\lambda=3.0$); c) a little girl with twin ponytails ($g=1.2g_s,
\protect\lambda=1.5$); d) a lady with luxuriant hair and shawl ($g=3.4g_s,
\protect\lambda=0.8$); e) an emperor penguin ($g=0.4g_s, \protect\lambda=2.0$%
); f) an outward-flying swallow in radar-invading mode ($g=0.22g_s, \protect%
\lambda=2.0$). For a better
visibility the spin amplitude is amplified by $\langle \sigma
_{z,x}\rangle^A$ with $A=1/4, 1/2,
1/3$ for (a-d), (e), (f) respectively. Here $j_E=5$ and
$\{n_Z, n_w, n_{ex}, n_{DK} \}$ =
$\{ 1,1,2,0 \}$ (a),
$\{ 3,-3,2,0 \}$ (b),
$\{ 5,-1,2,0 \}$ (c),
$\{ 4,4,2,2 \}$ (d),
$\{ 4,0,4,2 \}$ (e),
$\{ 3,3,0,4 \}$ (f).}
\label{fig-Cartoons}
\end{figure*}

\begin{figure*}[t]
\includegraphics[width=2.0\columnwidth]{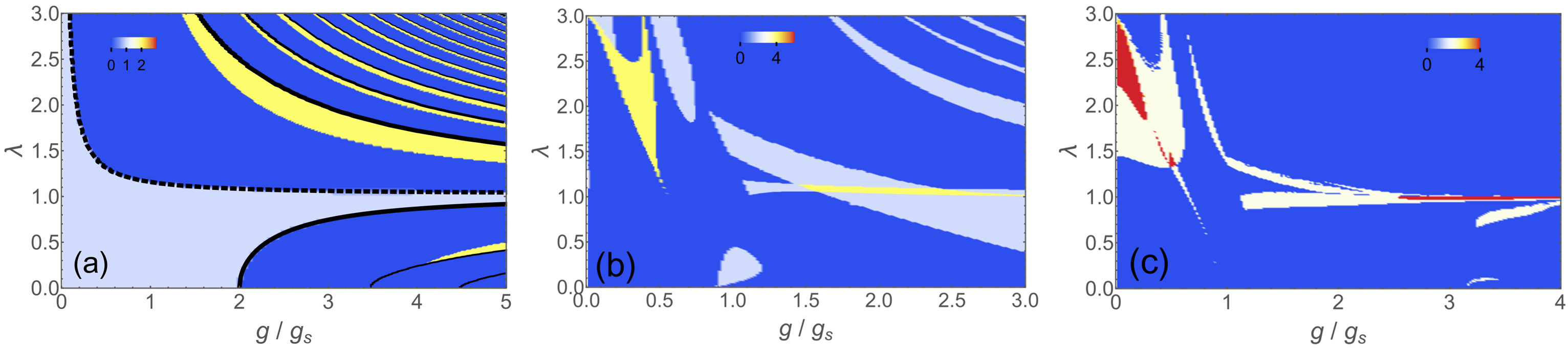}
\caption{a) Phase diagram of $n_{ex}$ in the ground state: Hidden small-spin-knot transitions. The solid and dashed lines mark
the previous conventional and unconventional TTs for comparison. b) Phase diagram of $%
n_{ex}$ for state $j_E=5$. c) Phase diagram of $n_{DK}$ for excited state $%
j_E=5$. }
\label{fig-Phase-nM-nSide}
\end{figure*}

\section{Hidden Transitions of Spin Knots}

\label{Sect-Hidden-Knots-Trans}

The topology of knot depends on what criterion one adopts. For a rope in
three-dimensional space, a knot with only one returning loop can be
continuously deformed into a straight line, such a knot topologically is
equal to a straight line. However in a two-dimensional plane, if one counts
the number of holes formed by loops of returning paths and classify the
topology by the corresponding fixed hole numbers, a knot is then topologically
different from a straight line. In this sense, spin knots reflect a
different level of topological information apart from the topology of
wave-function nodes on an axis and spin winding around the origin. In this
section we shall explore more hidden transitions induced by spin knots.

\subsection{Small (Scallop) Knot Transitions in Ground State}

As mentioned in Section \ref{Sect-Bridge-Knot-TopTrans}, \textit{small spin
knots} are formed around $\langle \sigma _{x}\left( x\right) \rangle $. We
term small knot not by its absolute size but by the characteristic that it
only covers either positive or negative $\langle \sigma _{x}\left( x\right)
\rangle $ axis but not across the $\langle \sigma _{z}\left( x\right)
\rangle $ axis as the big (bridge) knots. Small spin knots do not break the
correspondence of wave-function node numbers and the spin winding numbers,
but they change the number of $\langle \sigma _{z}\left( x\right) \rangle $
zeros. We show a small knot in \textbf{Figure} \ref{fig-Cartoons}a, where
although the knot appears like a biggest hat of chef the knot does not span
or bend to cross the $\langle \sigma _{z}\left( x\right) \rangle $ axis. The
knot profile in Figure \ref{fig-Cartoons}a also reminds us of the exotic
shark scalloped hammerhead, while the small knot can also be really small as
a pearl in a scallop (e.g. in Figure \ref{fig-nodes-winding-2}f). The
feature of scallop shape here most distinguished from the big bridge knot is
that scallop never has bridge piers. So we also call a small knot \textit{%
scallop knot} if terming a group of knots by size is not a good way.

At this point it should be mentioned that the small (scallop) knot can also be formed by the two infinity
ends which are regarded as connected, as mentioned in Section \ref{sec-nw-Algeb-Formu}, which can be
tracked down by the condition $|m(n_{Z}^{(+,-)})|>1$. We show an example in Figure \ref{fig-Cartoons}b
which looks like a sumo wrestler. As one may recognizes the two arms are starting from the origin dot, which corresponds to infinity in $x$ space, and form a small knot at the neck position.

The node sorting and algebraic formulation for spin windings in last section
inspires a way to monitor the small spin knots quantitatively. We propose
the following quantity%
\begin{equation}
n_{ex}=\sum_{i=1}^{n_{Z}^{(+,-)}}\left(  m(i) -1\right),  \label{n-Knots}
\end{equation}%
which actually denotes the \textit{extra} zero number of $\langle \sigma
_{z}\left( x\right) \rangle $ on $\langle \sigma _{x}\left( x\right) \rangle
$ axis formed by the returning of the spin trajectory apart from the first
time going across the $\langle \sigma _{x}\left( x\right) \rangle $ axis.
Note that not only the knotless winding is not included in $n_{ex}$ but also
the bridge knot is excluded, thus $n_{ex}$ can make a distinguishment from
these cases and extract a deeper topological information apart from $n_{Z}$
and $n_{aw}$. A vanishing $n_{ex}$ guarantees no small knots, while $%
n_{ex}=2 $ means the existence of a small spin knot.
An odd number of $n_{ex}$ indicates both vanishing $n_Z$ and $n_w$, e.g. as in Figure \ref{fig-phase-antiWinding}c where the blue island around $g=3.3g_{\rm s}, \lambda=0.1$ where $n_{ex}=3$. Although $n_{ex}=1$ corresponds to the topologically trivial case as in Figure \ref{fig-nodes-winding-1}a a larger odd number will mean spin knots despite of no effective winding.

Indeed, although there are no anti-winding large knots in the ground state, small-knot transition can still occur. We show $n_{ex}$ of the
ground state in \textbf{Figure} \ref{fig-Phase-nM-nSide}a. As one sees $%
n_{ex}$ is zero (blue) in most regions while, apart from $n_{ex}=1$ in the no-node region (light blue), another finite value ($n_{ex}=2$)
appears in slim yellow belts adjacent to the conventional topological boundaries
with level closing (black solid lines). These belt regions show the presence of a
small spin knot, which occurs mostly in $\lambda >1$ regime and turns to be
thin silvers in regime of larger $\lambda $ and $g$. It should be mentioned
here that, unlike the chef hat in Figure \ref{fig-Cartoons}a, the spin knot in the ground
state is really small as in Figure \ref{fig-nodes-winding-2}f. These new
transitions emerge in gapped situation, thus also being unconventional
TTs in addition to the previous unconventional boundary (black dashed
line) found from transition of the wave-function nodes in the gapped phase.

To sum up, now for the ground state we have two kinds of unconventional
TTs in different levels, one from the node transition,
the other from the small-spin-knot transition.

\subsection{Huge (Hug) Knots in Excited states}

The formulation of $n_{ex}$ in (\ref{n-Knots}) helps us to reveal knots
beyond the small and big ones. In fact, apart from the odd-number case which has no effective
winding as mentioned below Equation (\ref{n-Knots}), an even number of $n_{ex}$ larger than $2$
is also possible and can occur in three cases: (i) there are more than one
small knots but each appears in different position section $%
[x_{Z,i}^{(+,-)},x_{Z,i+1}^{(+,-)}]$; (ii) more than one small knots within
a same section $[x_{Z,i}^{(+,-)},x_{Z,i+1}^{(+,-)}]$, (iii) there are some
\textit{huge knots} that span over positive and negative $\langle \sigma
_{x}\left( x\right) \rangle $ axes through $\langle \sigma _{z}\left(
x\right) \rangle $ axis. Case (i) belongs to the afore-mentioned normal
small knots. Case (ii) is rare in low-lying states. Case (iii) is more
often. We give an example of case (iii) in Figure \ref{fig-Cartoons}e where
the spin trajectory draws a shape of penguin. One sees that two huge knots
are hugging the origin to build the head, body and leg parts of the penguin.
With this hugging shape we also call the huge knots \textit{hug knots}.

\subsection{Diagonal (Dipterus) Knots in Excited States}

The huge (Hug) knots revealed by $n_{ex}$ again leads to our finding of
another kind of knots which are distinguished from all the afore-addressed
small, big and huge knots. In the penguin-like case we also see the two fins
of the penguin which form two narrow spin knots. These knots are not located
on the $\langle \sigma _{x}\left( x\right) \rangle $ or $\langle \sigma
_{z}\left( x\right) \rangle $ axis as the small, big and huge knots but have
orientations between the axes, which we call \textit{diagonal Knots.} Note
that, except the condition of being away from the axes where the spin zeros
are located, the diagonal knots are not directly associated with the zeros
of $\langle \sigma _{x}\left( x\right) \rangle $ or $\langle \sigma
_{z}\left( x\right) \rangle$. Nevertheless, they can be tracked down by the
equal-spin condition
\begin{equation}
\langle \sigma _{z}\left( x_{1}\right) \rangle +i\langle \sigma _{x}\left(
x_{1}\right) \rangle =\langle \sigma _{z}\left( x_{2}\right) \rangle
+i\langle \sigma _{x}\left( x_{2}\right) \rangle ,\quad
\end{equation}%
of two different points $x_{1},x_{2}$ between a $\langle \sigma _{x}\left(
x\right) \rangle $ zero and a neighboring $\langle \sigma _{z}\left(
x\right) \rangle $ zero. We can also have more pairs of diagonal knots in a
same state, as in Figure \ref{fig-Cartoons}f where the spin winding draws a
shape like an outward-flying swallow. As in radar-invading mode we don't see
much of the wide wings which appear to be two narrow diagonal knots.
However, the body and tail of the swallow are not in a good stealth and we
see the forked tail that appears to be another two diagonal knots.
Considering the above penguin and swallow examples, we also call diagonal
knots \textit{dipterus knots}.

\subsection{Composite Knots in Excited States}

We have introduced different kinds of knots in different states. One can
also have different kinds of knots in a same excited state. For an example,
Figure \ref{fig-Cartoons}c shows a spin winding with a little-girl-alike
shape. The twin ponytails from the top are two big knots, while the shoulder
is a small knot around negative-$\langle \sigma _{x}\left( x\right) \rangle $
axis. These two kinds of knots appear in a same state but well separated in
the evolution with respect to $x$. The state in Figure \ref{fig-Cartoons}e
with a penguin shape has two huge knots and two diagonal knots at the same
time. Although the huge knots have an overwhelming size, basically the huge
knots and the diagonal knots are not mixing up with each other. Still, there
can be composite knots by mixed knots. We show a case in Figure \ref%
{fig-Cartoons}d which abstractly looks like a lady with luxuriant hair. Note
the hair style is changed from twin ponytails in Figure \ref{fig-Cartoons}c
to thick layered Bob, there are no anti-winding-node large knots now. Of course one can also have braids formed by
diagonal knots if moving a bit to around $\lambda =1$, we don't show the figures lest some one should think we are opening a hair salon.
Here in Figure \ref{fig-Cartoons}d
the shoulder changes to be a shawl-like shape which actually is composed of
one small knot around negative-$\langle \sigma _{x}\left( x\right) \rangle $
axis and two diagonal knots. Note the diagonal knots are formed within the
small knot, thus turning out to be a composite knot.
If we look back at
Figure \ref{fig-Cartoons}c with the little-girl shape, the left ponytail,
the shoulder and the right ponytail together actually form a giant composite
knot which meanders in a complicated returning path with canceling in
winding and several times of switching back and braiding.

\subsection{A Full Topological Identity Encoding Including Diagonal Spin Knots}

Since the newly found diagonal knots are not directly associated with the nodes of wave function or equivalently the zeros of the spin winding, we can assign a distinguished number, e.g. 5, to a diagonal knot in the topological code of eigenstate as mentioned in Section \ref{sec-Topo-Id}. Thus, the topological code now includes three angles or levels of topological information:
(i) the digits 1,3 represent the nodes of ${\psi }_{\pm }(x) $, i.e., the $\langle \sigma _{x}(x) \rangle $ zeros on positive-,negative-$\langle \sigma _{z}(x) \rangle $ axes in $\langle \sigma _{z}\left( x\right) \rangle $-$\langle \sigma
_{x}\left( x\right) \rangle $ plane;
(ii) the digits 2,4 denote the nodes of $\widetilde{\psi }_{\pm }\left( x\right) $, i.e, the $\langle \sigma _{z}(x) \rangle $ zeros on positive-,negative-$\langle \sigma _{x} (x) \rangle $ axes; (iii) the digit 5 marks a diagonal spin knot. For an example, Figure \ref{fig-Cartoons}e has two diagonal knots on the infinity sides which contribute two 5s in the two ends of the corresponding topological code $5 12 34432 14412 35 \widetilde{4}$. The digital order of the code reflects the detailed topological structure. Then, such a code encodes a full topological information of the eigenstate.

\subsection{Phase Diagrams of Extra Zeros and Diagonal Knots}

Besides the small-knot phase diagram of the ground state in Figure \ref%
{fig-Phase-nM-nSide}a, for the excited states we show the phase diagrams of
the extra zero number $n_{ex}$ on $\langle \sigma _{x}\left( x\right)
\rangle $ axis and diagonal-knot number $n_{DK}$ in Figure \ref%
{fig-Phase-nM-nSide}b,c. We see that small knots not only appear in large-$%
g,\lambda $ regime similarly to the ground state, but also emerge in
intermediate- and small-$\lambda $ regimes for a coupling $g\gtrsim g_{%
\mathrm{s}}$. Huge knots are formed in the large-$\lambda $ but small-$g$
regime. Diagonal knots show up not only in large-$\lambda $ and small-$g$
regime also also possible around $\lambda =1$ for $g\gtrsim g_{\mathrm{s}}$.
Simultaneous appearances of different kinds of knots can be found in the
overlapping regime of these phase diagrams as well as those in Figure \ref%
{fig-phase-antiWinding}e.

\subsection{Anisotropic QRM as a Born Abstract Artist}

As in the above descriptions for different knots we are surprised to find
that the sophisticated drawings of the spin windings in Figure \ref%
{fig-Cartoons}a-f really have amazing spiritual similarity with a chef in a
biggest hat and with beard, a sumo wrestler, a little girl with twin
ponytails, a lady with luxuriant hair and shawl, an emperor penguin, and an
outward-flying swallow in radar-invading mode, respectively. One may wonder
about Figure \ref{fig-phase-antiWinding}b to which we have not paid attention before as a painting but now we see a beard judge in wigs or one may say a pilot in helmet and oxygen mask.
It should be mentioned that, except for
the finite numbers of $n_{ex}$, $n_{DK}$ in Figure \ref{fig-Phase-nM-nSide}%
b,c, these figure illustrations are chosen not deliberately, as one can see
from the corresponding parameter digits which are not fine-tuned. In fact,
we simply pick up a point in the $n_{ex}$, $n_{DK}$ phase diagrams quite
randomly and it just turns out to be an artistic work. One may question
about the earlier works in Figures \ref{fig-nodes-winding-1},\ref%
{fig-nodes-winding-2} which were not in the art gallery. Well ... what we
can interpret first is: Figure \ref{fig-nodes-winding-2}f may be some star
gate or teleportation device and Figure \ref{fig-nodes-winding-1}e is the
corresponding exit tunnel. We are not sure whether or not the teleportation is done
with full fidelity for which one has to pray before going out of the exit.
For the other panels one needs to put upside down: In Figure {%
\rotatebox{180}{\ref{fig-nodes-winding-2}} panel {\rotatebox{180}{(e)}} is a
butterfly spy (probably still in experimental stage), panel {\rotatebox{180}{(d)}} is a devilbat associated from devilfish or a Carnival cat
mask; In Figure {\rotatebox{180}{\ref{fig-nodes-winding-1}} panel {%
\rotatebox{180}{(f)}} is a lotus flower, and finally panel {%
\rotatebox{180}{(d)}} is without doubt the most abstract portrait in the
world of an extraterrestrial being who even seems to be quite amiable. }}

Amazed by the painting sophistication, abstraction, imagination,
spirituality, quality, variety, prolificity, and so forth, we cannot help
drawing the conclusion that the anisotropic QRM is a born topological
abstract artist~\cite{AbstractArt} whose talent apparently has been buried
in the numerous findings about the QRM and its extensions including symmetry,%
\cite%
{Braak2019Symmetry,HiddenSymMangazeev2021,HiddenSymLi2021,HiddenSymBustos2021}
various patterns of symmetry breaking,\cite%
{Ying2020-nonlinear-bias,Ying-2018-arxiv,Ying-2021-AQT} few-body QPTs,\cite%
{Liu2021AQT,Ashhab2013,Ying2015,Hwang2015PRL,Ying2020-nonlinear-bias,Ying-2021-AQT,LiuM2017PRL,Hwang2016PRL,Ying-gapped-top,Ying-2018-arxiv,Ying-Stark-top}
multicriticalities and multiple points,\cite%
{Ying2020-nonlinear-bias,Ying-2021-AQT,Ying-gapped-top,Ying-Stark-top}
universality classification,\cite%
{Hwang2015PRL,LiuM2017PRL,Irish2017,Ying-2021-AQT,Ying-Stark-top} spectral
collapse,\cite%
{Felicetti2015-TwoPhotonProcess,e-collpase-Garbe-2017,e-collpase-Duan-2016,CongLei2019,Rico2020}
photon blockade effect,\cite%
{Boite2016-Photon-Blockade,Ridolfo2012-Photon-Blockade} spectral conical
intersections,\cite{Li2020conical} classical-quantum correspondence,\cite%
{Irish-class-quan-corresp} single-qubit topological phase transitions,\cite%
{Ying-2021-AQT,Ying-gapped-top,Ying-Stark-top} and so on. Since the anisotropic QRM might be the
Picasso~\cite{Picasso} of physical models,  maybe now it is
the moment to give some voice to art.

\section{Conclusions and Discussions}

\label{Sect-Conclusions}

With the exploration of underlying TTs via a thorough study on wave-function
nodes and spin windings, we have shed a new light on the energy spectrum of
the anisotropic QRM which is the fundamental model of light-matter
interactions with indispensable counter-rotating terms in ultra-strong
couplings.

On the one hand, by tracking the variation of node numbers in the eigen wave
functions, besides the conventional TTs at level crossings we have revealed
emerging unconventional TTs without gap closing or parity reversal
underlying the level anticrossings. Such level-anticrossing-connected
unconventional TTs do not occur in the ground state due to the absence of
level anticrossing, which accounts for the pureness of the ground state
topological phase diagram. Apart from the level-anticrossing-connected
unconventional TTs, a particular unconventional TT without level
anticrossing is also found and turns out to be universal for the ground
state and excited states. This particular transition have several advantages
in potential applications for quantum sensors or devices, such as large gap
situation, without limitation to ground state, and being applicable for all
coupling regimes.

On the other hand, we find that the wave-function nodes have a
correspondence to the zeros of spin windings, which endows the nodes a more
explicit topological character and provides a physical support for
single-qubit topological phase transitions. Thus, the node number or spin
winding number can be used as quantum topological numbers along with the
parity to characterize the quantum states especially in the ground states.
In such a topological classification, the clockwise/counterclockwise
direction of the spin winding adds another quantum feature to further
distinguish the various quantum states of light-matter interactions.

Moreover, when the wave-function node number is corresponding to the spin
winding number in the ground state, another kind of TTs arise in the excited
states with unmatched node number and winding number due to the emerging
anti-winding spin knots. Hidden transitions of small knots are found for the
ground state, while in excites states transitions of different spin knots
emerge including small (scallop), big (bridge), huge (hug) and diagonal
(dipterus) ones. The analysis on the node sorting and ordering leads us to an
algebraic formulation of the spin winding number which originally is in
integral form, building a bridge of geometric topology and algebraic
topology in a physical way concerning the wave function and the spin
windings.

Since the spin is a physical quantity, the topological information
originally encoded in the topological structure of the wave function now can
be decoded by the spin texture. Such a physical decoding not only turns the
topological information to be detectable, but also might provide possibility
for designing topological quantum devices or sensors. In such a perspective,
both the conventional TTs and unconventional ones addressed in the present
work might be have some potential applications as both gap-closing~\cite%
{Garbe2021-Metrology} and gapped~\cite{Ying2022-Metrology} situations are
applicable, while the unconventional TTs might have some more advantage in
avoiding the detrimental slowing-down effect close to transitions.~\cite%
{Ying2022-Metrology}

It should be noted that the anisotropic QRM considered in the present work
is a realistic model which can be implemented in superconducting circuits
with possible access to ultra-strong~\cite%
{Diaz2019RevModPhy,Wallraff2004,Gunter2009,Niemczyk2010,Peropadre2010,FornDiaz2017,Forn-Diaz2010,Scalari2012,Xiang2013,Yoshihara2017NatPhys,Kockum2017}
and even deep-strong couplings,\cite%
{Yoshihara2017NatPhys,Bayer2017DeepStrong} while the interaction anisotropy
is also highly tunable.~\cite{PRX-Xie-Anistropy,Forn-Diaz2010,Yimin2018}
Experimentally in such circuit systems\cite%
{flux-qubit-Mooij-1999,Bertet2005mixedModel,you024532} the effective
position $x$ (momentum $p$) can be simulated by the flux (charge) of
Josephson junctions which can be continuously tuned and the spin texture
might be measured by interference devices and magnetometer.\cite{you024532}
We speculate tests or applications of our results might be feasible in these
platforms. On the other hand, our analysis and the gained insight might also be helpful or relevant for some other systems as our model
shares some similarity with those in nanowires,\cite{Nagasawa2013Rings,Ying2016Ellipse,Ying2017EllipseSC,Ying2020PRR,Gentile2022NatElec} cold atoms~\cite{Li2012PRL,LinRashbaBECExp2011} and relativistic systems.~\cite{Bermudez2007}

As a final remark, our finding shows that the anisotropic QRM might be the
Picasso~\cite{Picasso} of physical model community. To this extent, one may
think art is starting to join the dialogue between mathematics and physics~\cite{Solano2011}
which was triggered by the milestone work of D. Braak~\cite{Braak2011} on the integrability of
the QRM as mentioned in the beginning of Introduction.

\section*{Acknowledgements}

This work was supported by the National Natural Science Foundation of China
(Grant No. 11974151).


\begin{thebibliography}{999}
\bibitem{AbstractArt} Before the dizzying painting factions including
impressionism, post-impressionism, expressionism, cubism, fauvism,
minimalism, abstractionism, modernism, surrealism, and so on and so forth,
we didn't find ourselves until coming across the word radical ``abstract"
which we always have to deal with in all scientific papers.
\bibitem{Picasso} Although in conventional classification of painting
factions P. Picasso is the founder of cubism and a main representative of
modernism, rather than abstractionism, we choose Picasso here as the iconic
figure for the reason that we as laymen heard his name more, for which we
hope the anisotropic QRM wouldn't mind.
\end{thebibliography}
\end{document}